\documentclass[12pt, a4paper]{article}
\usepackage{amsfonts, amssymb}
\usepackage{graphicx}
\usepackage{latexsym,amsmath,color,array}

\usepackage{natbib}
\usepackage{hyperref}
\usepackage{enumitem}
\usepackage{bbding}
\usepackage{amssymb}
\usepackage{amsmath}
\usepackage{graphicx}
\usepackage{amsmath}
\usepackage{bbm}
\usepackage{mathtools}
 \usepackage{color}
 \usepackage{array}
 \usepackage{subcaption}
\usepackage{caption}
 \usepackage{multirow}
 \usepackage[dvipsnames]{xcolor}
 \usepackage{makecell}
 \usepackage{colortbl}
 \usepackage{algorithm}
 \usepackage{url}
 \usepackage{tikz}
\usetikzlibrary{positioning}
\usetikzlibrary{decorations.pathreplacing,calligraphy}
\usetikzlibrary{external}
\hypersetup{
    hidelinks = true
}
\usepackage{url}

  \usepackage{booktabs}
 \usepackage{threeparttable}
 
  \usepackage{subcaption}
\DeclareCaptionLabelFormat{andfigure}{#1~#2  \&  \figurename~\thefigure}

\def\1{\mathbb{I}}

\newcounter{appen}[section]

\begin{document}

\title{A Statistical-Modelling Approach to Feedforward Neural Network Model Selection}
\author{Andrew McInerney\footnote{Department of Mathematics and Statistics, University of Limerick; andrew.mcinerney@ul.ie} \hspace{3cm}
Kevin Burke\footnote{Department of Mathematics and Statistics, University of Limerick; kevin.burke@ul.ie}}
\date{\today}

\maketitle

\begin{abstract}
Feedforward neural networks (FNNs) can be viewed as non-linear regression models, where covariates enter the model through a combination of weighted summations and non-linear functions.
Although these models have some similarities to the approaches used within statistical modelling, the majority of neural network research has been conducted outside of the field of statistics.
This has resulted in a lack of statistically-based methodology, and, in particular, there has been little emphasis on model parsimony.
Determining the input layer structure is analogous to variable selection, while the structure for the hidden layer relates to model complexity.
In practice, neural network model selection is often carried out by comparing models using out-of-sample performance.
However, in contrast, the construction of an associated likelihood function opens the door to information-criteria-based variable and architecture selection.
A novel model selection method, which performs both input- and hidden-node selection, is proposed using the Bayesian information criterion (BIC) for FNNs.
The choice of BIC over out-of-sample performance as the model selection objective function leads to an increased probability of recovering the true model, while parsimoniously achieving favourable out-of-sample performance.
Simulation studies are used to evaluate and justify the proposed method, and applications on real data are investigated.

\smallskip

{\bf Keywords.} Neural networks; Model selection; Variable selection; Information criteria.

\end{abstract}

\qquad

\newpage
\section{Introduction}
Neural networks are a popular class of machine-learning models, which pervade modern society through their use in many artificial-intelligence-based systems \citep{lecun2015deep}. 
Their success can be attributed to their predictive performance in an array of complex problems \citep{state_of_nn2018}.
Recently, neural networks have been used to perform  tasks such as natural language processing \citep{goldberg2016primer}, anomaly detection \citep{pang2021deep}, and image recognition \citep{voulodimos2018deep}.
Feedforward neural networks (FNNs), which are a particular type of neural network, can be viewed as non-linear regression models, and have some similarities to statistical modelling approaches (e.g., covariates enter the model through a weighted summation, and the estimation of the weights for an FNN is equivalent to the calculation of a vector-valued statistic) \citep{ripley1994neural, white1989learning}.
Despite early interest from the statistical community \citep{white1989learning, ripley1993statistical, cheng1994neural}, the majority of neural network research has been conducted outside of the field of statistics \citep{breiman2001statistical, hooker2021bridging}.
Given this, there is a general lack of statistically-based methods, such as model and variable selection, which focus on developing parsimonious models.

Typically, the primary focus when implementing a neural network centres on model predictivity (rather than parsimony); the models are viewed as `black-boxes' whose complexity is not of great concern \citep{efron2020prediction}.
It is perhaps not surprising, therefore, that there is a tendency for neural networks to be highly over-parameterised, miscalibrated, and unstable \citep{SUN2022109246}.
Nevertheless, FNNs can capture more complex covariate effects than is typical within popular (linear/additive) statistical models.
Consequently, there has been renewed interest in merging statistical models and neural networks, for example, in the context of flexible distributional regression \citep{rugamer2020semistructured} and mixed modelling \citep{tran2020bayesian}.
However, statistically-based model selection procedures are required to increase the utility of the FNN within the statistician's toolbox.

Traditional statistical modelling is concerned with developing parsimonious models, as it is crucial for the efficient estimation of covariate effects and significance testing \citep{efron2020prediction}.
Indeed, model selection (which includes variable selection) is one of the fundamental problems of statistical modelling \citep{fisher1922mathematical}.
It involves choosing the ``best" model, from a range of candidate models, by trading pure data fit against model complexity \citep{anderson2004model}.
As such, there has been a substantial amount of research on model and variable selection \citep{miller2002subset}.
As noted by \citet{heinze2018variable}, typical approaches include significance testing combined with forward selection or backward elimination (or a combination thereof); information criteria such as AIC or BIC \citep{Akaike1998, schwarz1978estimating, anderson2004model}; and penalised likelihood such as LASSO \citep{tibshirani1996regression, fan2010selective}.

In machine learning, due to the focus on model predictivity, relatively less emphasis is placed on finding a model that strikes a balance between complexity and fit.
Looking at FNNs in particular, the number of hidden nodes is usually treated as a tunable hyperparameter \citep{bishop1995neural_9, pontes2016design}.
Input-node selection is not as common, as the usual consensus when fitting FNNs appears to be similar to the early opinion of \citet{breiman2001statistical}: ``the more predictor variables, the more information".
However, there are some approaches in this direction, and a survey of variable selection techniques in machine learning can be found in \citet{chandrashekar2014survey}.
Nevertheless, typically, the optimal model is usually determined based on its predictive performance, such as out-of-sample mean squared error, which can be calculated on a validation data set. 
Unlike an information criteria, out-of-sample performance does not directly take account of model complexity.

When framing an FNN statistically, there are several motivating reasons for a model selection procedure that aims to obtain a parsimonious model.
For example, the estimation of parameters in a larger-than-required model results in a loss in model efficiency, which, in turn, leads to less precise estimates.
Input-node selection, which is often ignored in the context of neural networks, can provide the practitioner with insights on the importance of covariates.
Instead, other feature importance measures are typically used such as the feature attribution methods described in \citet{koenen2024interpreting}.
Furthermore, eliminating irrelevant covariates can result in cheaper models by reducing potential costs associated with data collection (e.g. financial, time, energy).
In this paper, we take a statistical-modelling view of neural network selection by assuming an underlying (normal) error distribution.
Doing so enables us to construct a likelihood function, and, hence, carry out information-criteria-based model selection, such as the BIC \citep{schwarz1978estimating}, naturally encapsulating the parsimony in the context of a neural network.
More specifically, we propose an algorithm that alternates between selecting the hidden layer complexity and the inputs with the objective of minimizing the BIC.
We have found, in practice, that this leads to more parsimonious neural network models than the more usual approach of minimizing out-of-sample error, while also not compromising the out-of-sample performance itself.

The remainder of this paper is structured as follows.
In Section \ref{sec: fnn}, we introduce the FNN model while linking it to a normal log-likelihood function.
Section \ref{sec: model_sel} motivates and details the proposed model selection procedure.
Simulation studies to investigate the performance of the proposed method, and to compare it to other approaches, are given in Section \ref{sec: sim}.
In Section \ref{sec: app_to_data}, we apply our method to real-data examples.
Finally, we conclude in Section \ref{sec: disc} with a discussion.

\section{Feedforward Neural Network}\label{sec: fnn}
\begin{figure}[h]
    \centering
    \includegraphics[width=7.5cm]{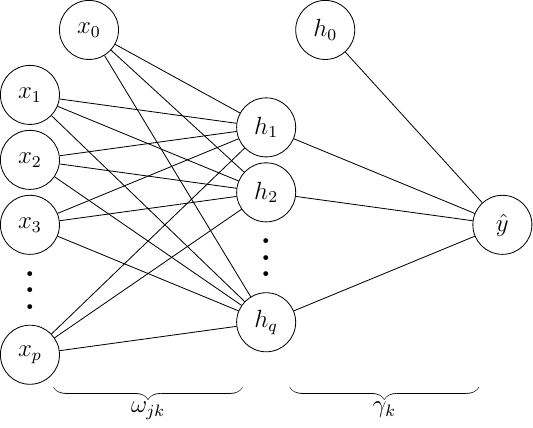}
    \caption{Neural network architecture with $p$ input nodes and $q$ hidden nodes.}
    \label{fig: nn_diagram}
\end{figure}

Let $y = (y_1,y_2,\dotsc,y_n) \in \mathbb{R}^n$ be the response variable of interest for a regression-based problem, where $n$ represents the number of observations. 
For the $i$th observation, $i=1,\dotsc,n$, let $x_i=(x_{1i},x_{2i},\dotsc,x_{pi})^T$ be a vector of $p$ covariates---the inputs to the neural network model.
We assume a model of the form
$y_i = \text{NN}(x_i) + \varepsilon_i,$
where $\varepsilon_i$ is a random error that we assume has a $N(0,\sigma^2)$ distribution, and $\text{NN}(\cdot)$ is a neural network,
\begin{equation}\label{eq: FNN}
  \text{NN}(x_i) = \gamma_0+\sum_{k=1}^q \gamma_k \phi \left( \sum_{j=0}^p \omega_{jk}x_{ji}\right).
\end{equation}
As we aim to frame FNNs as an alternative to other statistical non-linear regression models (i.e., used on small-to-medium sized tabular data sets relative to the much larger data sets seen more broadly in machine learning), and due to the universal approximation theorem \citep{cybenko1989approximation,hornik1989multilayer}, we are restricting our attention to FNNs with a single-hidden layer.
The parameters in Equation~\ref{eq: FNN} are as follows: $\omega_{0k}$, the intercept term associated with the $k$th hidden node; $\omega_{jk}$, the weight that connects the $j$th input node to the $k$th hidden node;
$\gamma_0$, the intercept term associated with the output node; and $\gamma_k$, the weight that connects the $k$th hidden node to the output node.
The function $\phi(\cdot)$ is the activation function for the hidden layer, which is often a logistic function.
The number of parameters in the neural network is given by 
$K = (p + 2)q + 1$.
A diagram of a neural network architecture with $p$ input nodes and $q$ hidden nodes is shown in Figure~\ref{fig: nn_diagram}.
In the diagram, $x_0 = 1$, $h_0 = 1$, and 
$h_k = \phi\left(\sum_{j=0}^p\omega_{jk}x_{ji}\right)$. 

Given our assumption that $\varepsilon_i \sim N(0, \sigma^2)$, we then make use of the log-likelihood function
\begin{equation}\label{eq: loglike}
  \ell(\theta)= -\frac{n}{2}\log(2\pi\sigma^2)-\frac{1}{2\sigma^2}\sum_{i=1}^n(y_i-\text{NN}(x_i))^2,
\end{equation}
where $\theta = (\omega_{01}, \ldots, \omega_{p1}, \ldots, \omega_{0q}, \ldots, \omega_{pq}, \gamma_0, \ldots, \gamma_q, \sigma^2)^T$.
We maximise this log-likelihood to obtain $\hat\theta$ but note that the estimates of the neural network parameters do not depend on the value of $\sigma^2$, i.e., the residual sum of squares, $\sum_{i=1}^n(y_i - \text{NN}(x_i))^2$, can be estimated to obtain the neural network parameters.
This is useful since standard neural network software (that minimises the residual sum of squares) such as \texttt{nnet} \citep{ripley2022nnet} can be used used to optimise the neural network followed by the estimation of $\sigma^2$ in a separate step.

The calculation of a log-likelihood function allows for the use of information criteria when selecting a given model, and in particular, the Bayesian information criterion (BIC) \citep{schwarz1978estimating},
$\text{BIC} = -2\ell(\hat{\theta}) + \log(n)(K+1)$,
where we have $K + 1$ parameters, i.e., the $K$ neural network parameters plus the variance parameter, $\sigma^2$.
An attractive property of the BIC is that it is ``dimension-consistent", i.e., the probability of selecting the ``true" model approaches one as sample size increases \citep{anderson2004model}.
It is important to note that other approaches for the calculation of the degrees of freedom exist \citep{murata1994nic, ye1998measuring}, but we find these do not penalise more complex models (with redundancies) heavily enough in the model selection context compared to using $K$ (see Appendix \ref{app: df}).

\section{Model Selection}\label{sec: model_sel}
To begin model selection, a set of candidate models must be considered.
For the input layer, we can have up to $p_{\text{max}}$ inputs, where $p_{\text{max}}$ is the maximum number of covariates being considered, and this is often the total number of covariates available in the data under study.
The input layer can contain any combination of these $p_{\text{max}}$ inputs.
For the hidden layer, we must specify a $q_{\text{max}}$ value, which is the maximum number of hidden nodes to be considered; this controls the maximum level of complexity of the candidate models.
We can then have between one and $q_{\text{max}}$ nodes in the hidden layer.
From a neural network selection perspective, we aim to select a subset of $p \le p_{\text{max}}$ covariates to enter the input layer and to build a hidden layer of $q \le q_{\text{max}}$ nodes to adapt to the required complexity.
To carry out these selections, we suggest a statistically-motivated procedure based on minimising the BIC, since it directly penalises complexity and is known to be selection consistent, i.e., BIC minimisation converges to the true model asymptotically.
In contrast, and more usually in machine learning applications, one could consider predictive performance, for example, the out-of-sample mean squared error.
We will also consider this approach but find that it leads to significantly more complex models than the use of BIC while only marginally improving predictive performance.
Whether one is aiming to minimise BIC or out-of-sample mean squared error, multiple initialisations of the neural network (from $n_\text{init}$ random vectors of parameters) are required to improve the chance of finding a global maximiser of the log-likelihood surface.

\subsection{Proposed Approach}\label{sec: approach}
\begin{figure}\centering
\includegraphics[width=13cm]{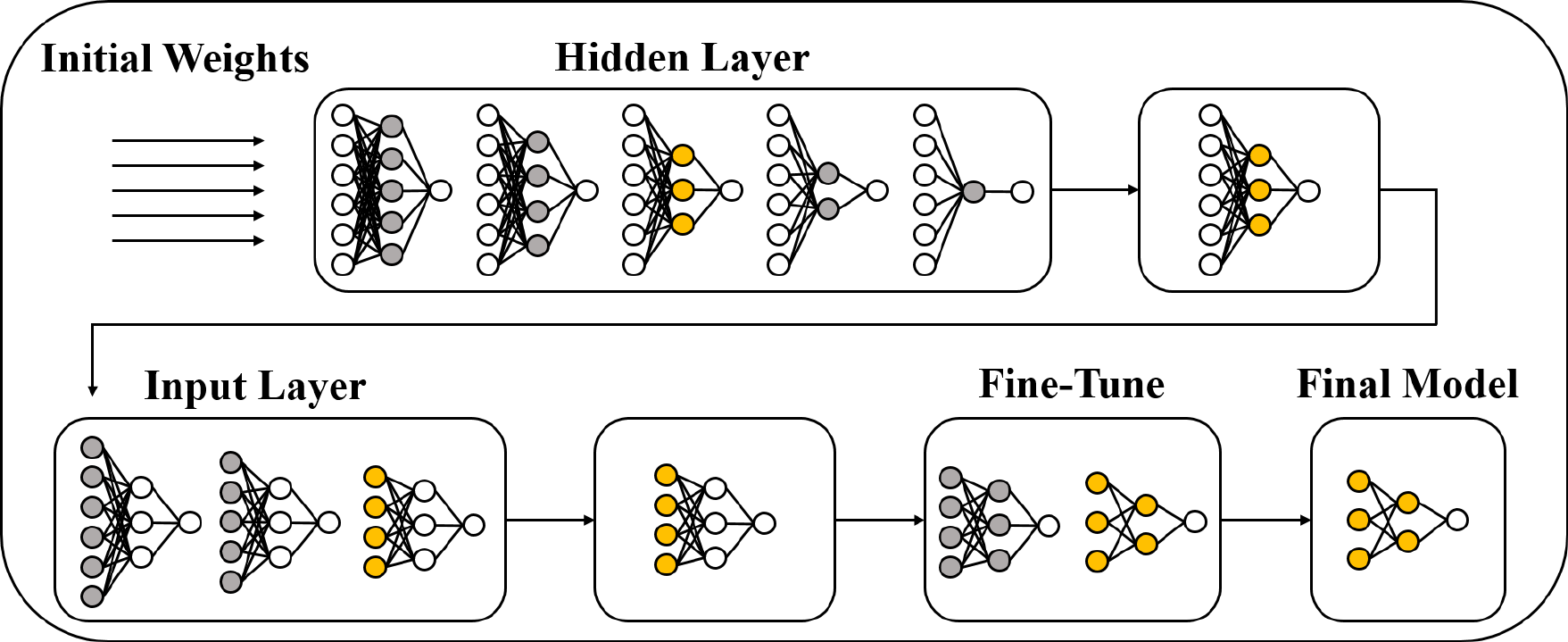}
\caption{Model selection schematic. Nodes coloured grey are being considered in current phase. Nodes coloured gold represent optimal nodes in that phase to be brought forward to the next phase.}
\label{mcinereny:schematic} 
\end{figure}

We propose a stepwise procedure that starts with a hidden-node selection phase followed by an input-node selection phase.
(We find that this ordering leads to improved model selection.)
This is, in turn, followed by a fine-tuning phase that alternates between the hidden and input layers for further improvements.
The proposed model selection procedure is detailed in Algorithm~\ref{alg: modelsel} (which relies on Algorithms \ref{alg: cand_model}--\ref{alg: input-node-sel}), and a schematic diagram is provided in Figure~\ref{mcinereny:schematic}.
It is also described at a high level in the following paragraphs.

The procedure (Algorithm \ref{alg: modelsel}) is initialised with the full set of input nodes, $\mathcal{X}_{\text{full}}$, the maximum number of hidden nodes being considered, $q_\text{max}$, and the number of initialisations, $n_{\text{init}}$, and, as mentioned, starts with a hidden-node selection phase (Algorithm \ref{alg: hidden-node-sel} with $Q = \{1, 2, \dotsc, q_{\text{max}} - 1\}$).
For each candidate model in this phase (i.e., models with $q \in \{1, \ldots, q_\text{max}\}$), the network optimiser is supplied with $n_\text{init}$ random vectors of initial parameters, the log-likelihood function is maximised at each of these vectors, and the overall maximiser is found (see Algorithm \ref{alg: cand_model}).
The reason for supplying the neural network with different vectors of initial parameters is due to the complex optimisation surface for neural networks that may contain several local maxima.
Thus, the use of a set of initial vectors (rather than just one) aims to increase the chance of finding the global maxima; of course, this cannot be guaranteed as is often the case in more complex statistical models.
Once all of the $q_\text{max}$ candidate models have been fitted, the hidden-node selection phase is concluded by selecting the one whose hidden structure (i.e., number of nodes, $q$) minimises the BIC.

\begin{algorithm}[t]
\small
\caption{Fit Candidate Model}
\label{alg: cand_model}

    \textbf{Input:} The set of input nodes, $\mathcal{X}$, the number of hidden nodes, $q$, and the number of initialisations, $n_{\text{init}}$.

        \begin{enumerate}[label=\textbf{\arabic*}., ref = \arabic*]
        
        \item Generate $n_{\text{init}}$ random initial weight vectors of size $K = (p + 2)q + 1$ with $p = |\mathcal{X}|$ and $|\cdot|$ is the cardinality of a set.
        \item Using a neural network optimiser, maximise the log-likelihood function (Equation~\ref{eq: loglike}) for each initialisation.
        \item Select the model with the maximum log-likelihood value as the candidate model.
        \item Calculate the associated BIC value.

    \end{enumerate}
    
    \textbf{Output:} A fitted neural network with its associated BIC value.
    
\end{algorithm}

\begin{algorithm}[ht!]
\small
\caption{Hidden-Node Selection}
\label{alg: hidden-node-sel}

\textbf{Input:} The set of input nodes currently included in the model, $\mathcal{X}$, the number of hidden nodes currently in the model, $q$, the set of hidden-layer structures being considered, $Q$, and the number of initialisations, $n_{\text{init}}$.

        \begin{enumerate}[label=\textbf{\arabic*}., ref = \arabic*]
        
        \item\textbf{For $\boldsymbol{k \text{ in } Q}$:} \\
        Perform Algorithm \ref{alg: cand_model} with the set $\mathcal{X}$ of input nodes, $k$ hidden nodes, and $n_{\text{init}}$ initialisations. \\
        \textbf{If $\boldsymbol{\text{BIC}(k) \leq \text{BIC}(q)}$:}\\
          Set $q = k$. 
        
    \end{enumerate}
    
        \textbf{Output:} The number of hidden nodes, $q$.
        
\end{algorithm}

\begin{algorithm}[ht!]
\caption{Input-Node Selection}
\small
\label{alg: input-node-sel}

\textbf{Input:} The set of all input nodes under consideration, $\mathcal{X}_{\text{full}}$, the set of input nodes currently included in the model, $\mathcal{X}$, the number of hidden nodes currently in the model, $q$, the limit on the number of iterations of the repeat step, $n_{\text{steps}}$, the number of initialisations, $n_{\text{init}}$, and this Algorithm covers the possibility of both dropping and adding input variables depending on whether Steps \ref{step: input-repeat}\ref{step: input-repeat-drop} and/or \ref{step: input-repeat}\ref{step: input-repeat-add} are applied.

        \begin{enumerate}[label=\textbf{\arabic*}., ref = \arabic*]
        
        \item 
        Set $i = 0$ and $\mathcal{X}_{\text{new}} = \mathcal{X}$.
        
        \item\textbf{Repeat:}\label{step: input-repeat}
        
        \begin{enumerate}[label=\textbf{(\alph*)}, ref = (\alph*)]
        
        \item\textbf{If drop inputs:}\label{step: input-repeat-drop}
        
        \begin{enumerate}[label=\textbf{\roman*.}, ref = \roman*.]
        
        \item\textbf{For $\boldsymbol{c \text{ in } \mathcal{X}}$:}\\
             Perform Algorithm \ref{alg: cand_model} with the set $\mathcal{X} \setminus \{c\}$ of input nodes, $q$ hidden nodes, and $n_{\text{init}}$ initialisations, where $\mathcal{X} \setminus \{c\}$ is the set $\mathcal{X}$ of input nodes with input node $c$ removed.
             
             \textbf{If $\boldsymbol{\text{BIC}(\mathcal{X} \setminus \{c\}) \leq \text{BIC}(\mathcal{X}_{\text{new}})}$:}\label{step: input-sel-drop}\\
         Set $\mathcal{X}_{\text{new}} = \mathcal{X} \setminus \{c\}$. 
        
        \end{enumerate}

         \item\textbf{If add inputs:}\label{step: input-repeat-add}
        
        \begin{enumerate}[label=\textbf{\roman*.}, ref = \roman*.]
        
        \item\textbf{For $\boldsymbol{c \text{ in } \mathcal{X}_{\text{full}} \setminus \mathcal{X}}$:}\\
             Perform Algorithm \ref{alg: cand_model} with the set $\mathcal{X} \cup \{c\}$ of input nodes, $q$ hidden nodes, and $n_{\text{init}}$ initialisations, where $\mathcal{X} \cup \{c\}$ is the set $\mathcal{X}$ of input nodes with input node $c$ added.
             
             \textbf{If $\boldsymbol{\text{BIC}(\mathcal{X} \cup \{c\}) \leq \text{BIC}(\mathcal{X_{\text{new}}})}$:}\label{step: input-sel-add}\\
         Set $\mathcal{X_{\text{new}}} = \mathcal{X} \cup \{c\}$. 
        
        \end{enumerate}
        
         \item\textbf{If $\boldsymbol{\mathcal{X} \neq \mathcal{X}_{\text{new}}}$:}\\
         Set $\mathcal{X} = \mathcal{X}_{\text{new}}$ and $i = i+1$.\\
         \textbf{Else:}\\
         End repeat.
         
          \item\textbf{If $\boldsymbol{i \geq n_{\text{steps}}}$:}\\
         End repeat.
 
        \end{enumerate}
        
    \end{enumerate}
    
     \textbf{Output:} The set of included input nodes, $\mathcal{X}$.
\end{algorithm}

\begin{algorithm}[ht!]
\caption{Model Selection}
\label{alg: modelsel}

\textbf{Input:} The set of all input nodes, $\mathcal{X}_{\text{full}} = \{x_1, x_2, \dotsc, x_{p_{\text{max}}}\}$, the maximum number of hidden nodes to be considered, $q_{\text{max}}$, and the number of initialisations, $n_{\text{init}}$.

        \begin{enumerate}[label=\textbf{\arabic*}., ref = \arabic*]
        
        \item \textbf{Hidden-Node Selection:}
        
        Perform Algorithm \ref{alg: hidden-node-sel} with $\mathcal{X} = \mathcal{X}_{\text{full}}$, $Q = \{1, 2, \dotsc, q_{\text{max}} - 1\}$, $q = q_{\text{max}}$, and $n_{\text{init}} = n_{\text{init}}$.

        \item \textbf{Input-Node Selection:}
        
        Perform Algorithm \ref{alg: input-node-sel} with $\mathcal{X}_{\text{full}} = \mathcal{X}_{\text{full}}$, $\mathcal{X}= \mathcal{X}_{\text{full}}$, $q = q$, $n_{\text{steps}} = p_\text{max}$, and $n_{\text{init}} = n_{\text{init}}$, applying only the ``drop inputs'' step.

        \item\textbf{Fine Tuning:} \label{step: finetune}
        
        \begin{itemize}
        
        \item\textbf{Repeat:}

         \begin{enumerate}[label=\textbf{(\alph*)}, ref = (\alph*)]
         
         \item\textbf{Hidden Layer:}\label{step: finetune-hidden}
         
         Perform Algorithm \ref{alg: hidden-node-sel} with $\mathcal{X} = \mathcal{X}$, $Q = \{q - 1, q + 1\}$, $q = q$, and $n_{\text{init}} = n_{\text{init}}$.
         
         \textbf{If Step~\ref{step: finetune}\ref{step: finetune-hidden} did not update the value of $\boldsymbol{q}$:} \\
        End repeat.$^*$
        
         \item\textbf{Input Layer:}\label{step: finetune-input}
         
         Perform Algorithm \ref{alg: input-node-sel} with $\mathcal{X}_{\text{full}} = \{x_1, x_2, \dotsc, x_{p_{\text{max}}}\}$, $\mathcal{X} = \mathcal{X}$, $q = q$, $n_{\text{steps}} = 1$, and $n_{\text{init}} = n_{\text{init}}$, applying both the ``drop inputs'' and ``add inputs'' steps.
         
         \textbf{If Step~\ref{step: finetune}\ref{step: finetune-input} did not update the value of $\boldsymbol{\mathcal{X}}$:} \\
        End repeat.$^*$
         
         \end{enumerate}
         \end{itemize}
    \end{enumerate}
     \textbf{Output:} The set of included input nodes, $\mathcal{X}$, and the number of hidden nodes, $q$.\\
         $\boldsymbol{^*}$ Note: The fine-tuning phase stops if either Step \ref{step: finetune}\ref{step: finetune-hidden} or Step \ref{step: finetune}\ref{step: finetune-input} does not find an improvement.
         This is to avoid either input-node or hidden-node selection being repeated under conditions previously considered.
\end{algorithm}

Once the hidden-node selection phase has concluded, the focus switches to the input layer (Algorithm \ref{alg: input-node-sel}); at this point, there are $p_\text{max}$ inputs (i.e., the set of input nodes currently included in the model is the set of all input nodes, $\mathcal{X} = \mathcal{X}_{\text{full}}$).
For the input-node selection phase, each input node is dropped in turn, with the aim of finding an input whose removal yields a lower BIC; as with the previous phase, random sets of initial parameters are used for each candidate model in the underlying likelihood optimisation.
If the removal of a given input node does yield a lower BIC value, then that input node is dropped from the model (and if two or more inputs result in a lower BIC, the one yielding the lower BIC is removed). 
This is repeated until no covariate, when removed from the model, results in a lower BIC, and, then, the set of included input nodes, $\mathcal{X}$, is returned.
(Thus, in this phase, Algorithm \ref{alg: input-node-sel} is applied with only the ``drop inputs'' step and $n_\text{steps} = p_\text{max}$.)

Both the hidden layer and covariate selection phases are backward elimination procedures.
Rather than stopping the algorithm after these two phases, we have found it fruitful to search for an improved model in a neighbourhood of the current ``best'' model by carrying out some further fine tuning.
This is done by considering the addition or removal of one hidden node (Algorithm \ref{alg: hidden-node-sel} with $Q = \{q - 1, q + 1\}$), then the further addition or removal of one input node (Algorithm \ref{alg: input-node-sel} with with both the ``drop inputs'' and ``add inputs'' steps and $n_{\text{steps}} = 1$), and these two steps are repeated alternately until no further adjustment decreases the BIC (see Step \ref{step: finetune} in Algorithm \ref{alg: modelsel}).
This fine-tuning stage is analogous to stepwise model selection with backward and forward steps.
Note that one could apply this alternating stepwise procedure from the offset, but we have found it to be significantly more computationally efficient to focus first on the hidden and input layers (separately and in that order) before moving to the stepwise phase.

The particular order of the model selection steps described above has been chosen in order to have a higher probability in recovering the ``true" model, and to have a lower computational cost (see Section \ref{sec: sim1} for a detailed simulation).
Note that choosing the \emph{set} of input nodes requires a more extensive search than choosing the \emph{number} of hidden nodes.
There are more candidate structures for the input layer as you can have any combination of the nodes.
Therefore, it is recommended to perform hidden-node selection first, to eliminate any redundant hidden nodes and decrease the number of parameters in the model, before performing input-node selection.

\section{Simulation Studies}\label{sec: sim}
In order to justify and evaluate the proposed model selection approach, three simulation studies are used:
\begin{itemize}
  \item \textbf{Simulation 1 (Section~\ref{sec: sim1}):} In our first simulation study, we investigate the effect of the ordering of the model selection steps to justify the procedure.
  This includes the effect of performing input-node and hidden-node selection phases first, the improvement of including a stepwise fine-tuning step, and the performance of a procedure that only carries out iterative stepwise steps (i.e., fine tuning from the offset).
  \item \textbf{Simulation 2 (Section~\ref{sec: sim2}):} The second simulation study compares the performance of using the BIC as the model selection objective function versus using AIC or out-of-sample mean squared error (OOS).

    \item \textbf{Simulation 3 (Section~\ref{sec: sim3}):} The third simulation study investigates the performance of the proposed model selection procedure in the case where the true data-generating process is not a neural network, but, rather is that of a linear-type regression model (albeit with non-linear and interaction terms). Here, we compare the performance of our procedure against classical linear-regression stepwise selection.
\end{itemize}

In the first two simulation studies, the response is generated from an FNN with known ``true" architecture.
The weights are generated so that there are three important inputs, $x_1, x_2, x_3$, with non-zero weights, and ten unimportant inputs, $x_4, \ldots, x_{13}$, with zero weights.
All input variables are independent and generated from a standard normal distribution and the error variance is 0.7 (but the results are similar when the inputs are correlated as shown in Appendix \ref{app: corr_data}).
The ``true" hidden layer consists of $q = 3$ hidden nodes, while we set our procedure to consider a maximum of $q_\text{max} = 10$ hidden nodes.
The weights of the neural network are held constant over all repetitions and are given by $(\omega_{01} = 1.40 , \omega_{11} = 4.35 , \omega_{21} = 3.22 , \omega_{31} = -2.43 , \omega_{02} = -2.89 , \omega_{12} = 4.28 , \omega_{22} = -3.27 , \omega_{32} = -2.30 , \omega_{03} = -1.90 , \omega_{13} = 4.49 , \omega_{23} = 3.24 , \omega_{33} = 2.46 , \gamma_{0} = 2.98 ,  \gamma_{1} = 2.37 ,   \gamma_{2} = 2.37 ,  \gamma_{3} = 2.47)^T$.
The metrics calculated to evaluate the performance of the model selection approach are the true negative rate (TNR) for the input nodes (i.e., the proportion of input nodes with true zero weights that are \emph{correctly} dropped from the model), the false discovery rate (FDR) for the input nodes (i.e., the proportion of input nodes with true zero weights that are \emph{incorrectly} included in the model), the average number of hidden nodes selected ($\bar{q}$),  the probability of choosing the correct set of inputs (PI), the probability of choosing the correct number of hidden nodes (PH), and the probability of choosing the overall true model (PT).
(All probabilities refer to the proportion of correct results from the 1,000 simulation replicates.)
In all simulation studies, we vary the sample size $n \in \{250, 500, 1000\}$ and carry out 1,000 replicates.
Our proposed model selection approach is implemented in our publicly available R package \texttt{selectnn} \citep{mcinerney2022selectnn}.
The neural network function used is \texttt{nnet}, which is available from the R package of the same name \citep{ripley2022nnet}.
(Note that we do not use a weight decay penalty when fitting the models, i.e., we set \texttt{decay = 0} within the \texttt{nnet} function.)

\subsection{Simulation 1: Model Selection Approach}\label{sec: sim1}
This simulation study aims to justify the approach of the proposed model selection procedure, i.e., a hidden-node phase, followed by an input-node phase, followed by a fine-tuning phase; here, we label this approach as H-I-F.
Some other possibilities would be: to start with the input-node phase (I-H-F), to stop the procedure without fine tuning (either H-I or I-H), or to only carry out fine-tuning from the beginning (F).
Descriptions of the considered model selection approaches are as follows (the proposed approach is highlighted in bold; round brackets indicate the reordering of the steps in Algorithm 4 required to achieve the approach): 
\begin{itemize}
    \item H-I: Hidden-node selection phase, followed by input-node selection phase (Step 1 $\rightarrow$ Step 2).
  \item  I-H: Input-node selection phase, followed by hidden-node selection phase (Step 2 $\rightarrow$ Step 1).
  \item \textbf{H-I-F}: \textbf{Hidden-node selection phase, followed by input-node selection phase, and then a fine-tuning phase (Step 1 $\rightarrow$ Step 2 $\rightarrow$ Step 3).} 
  \item I-H-F: Input-node selection phase, followed by hidden-node selection phase, and then a fine-tuning phase (Step 2 $\rightarrow$ Step 1 $\rightarrow$ Step 3). 
  \item F: Fine-tuning phase only (Step 3). 
\end{itemize}

\begin{figure}[t]
    \centering
    \includegraphics[width=13cm]{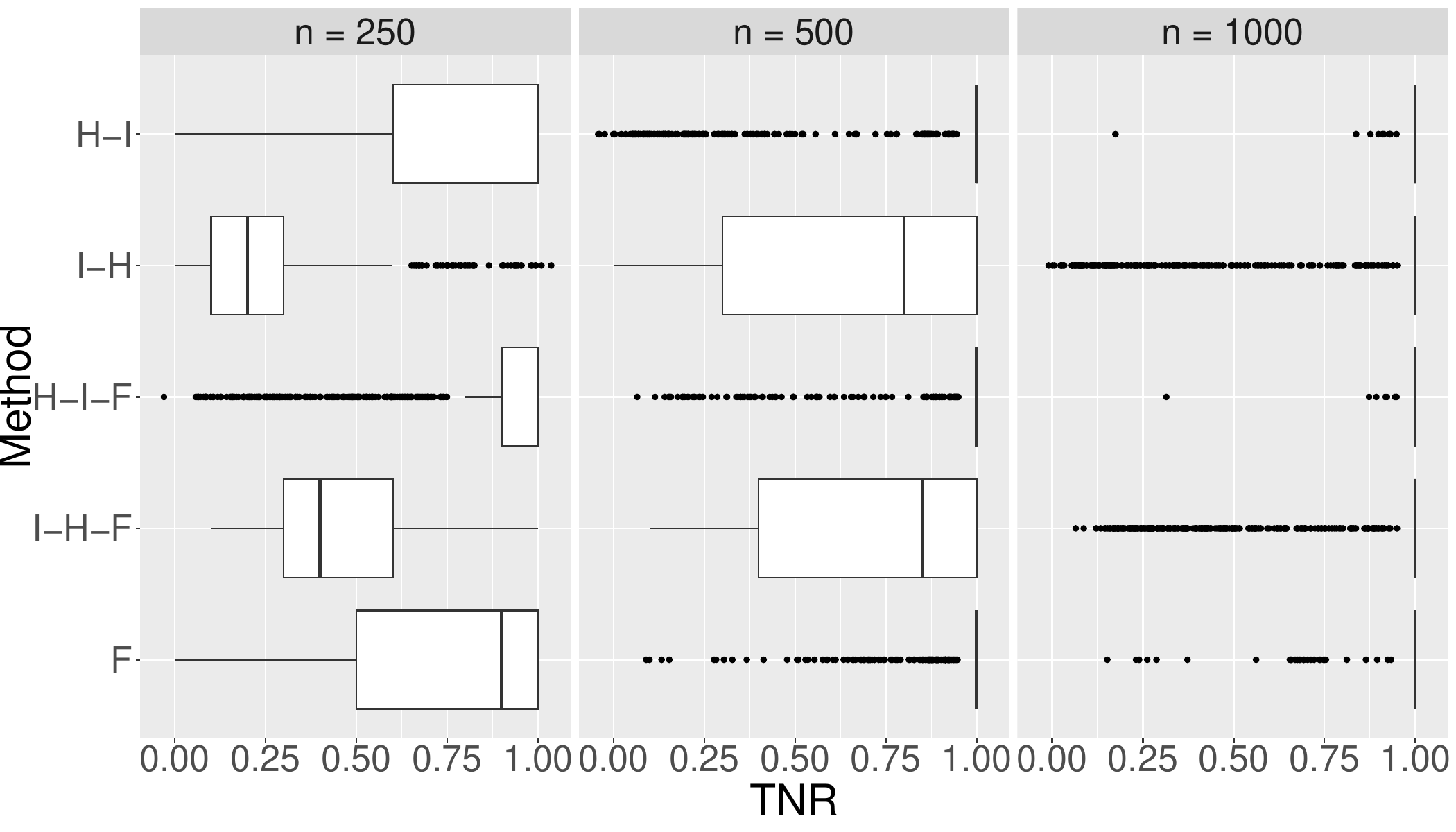}
    \caption{Simulation 1: boxplots for TNR (the true negative rate for the input variables) for each method by sample size.}
    \label{fig: sim1_boxplot_C}
\end{figure}

\begin{figure}[h!]
    \centering
    \includegraphics[width=13cm]{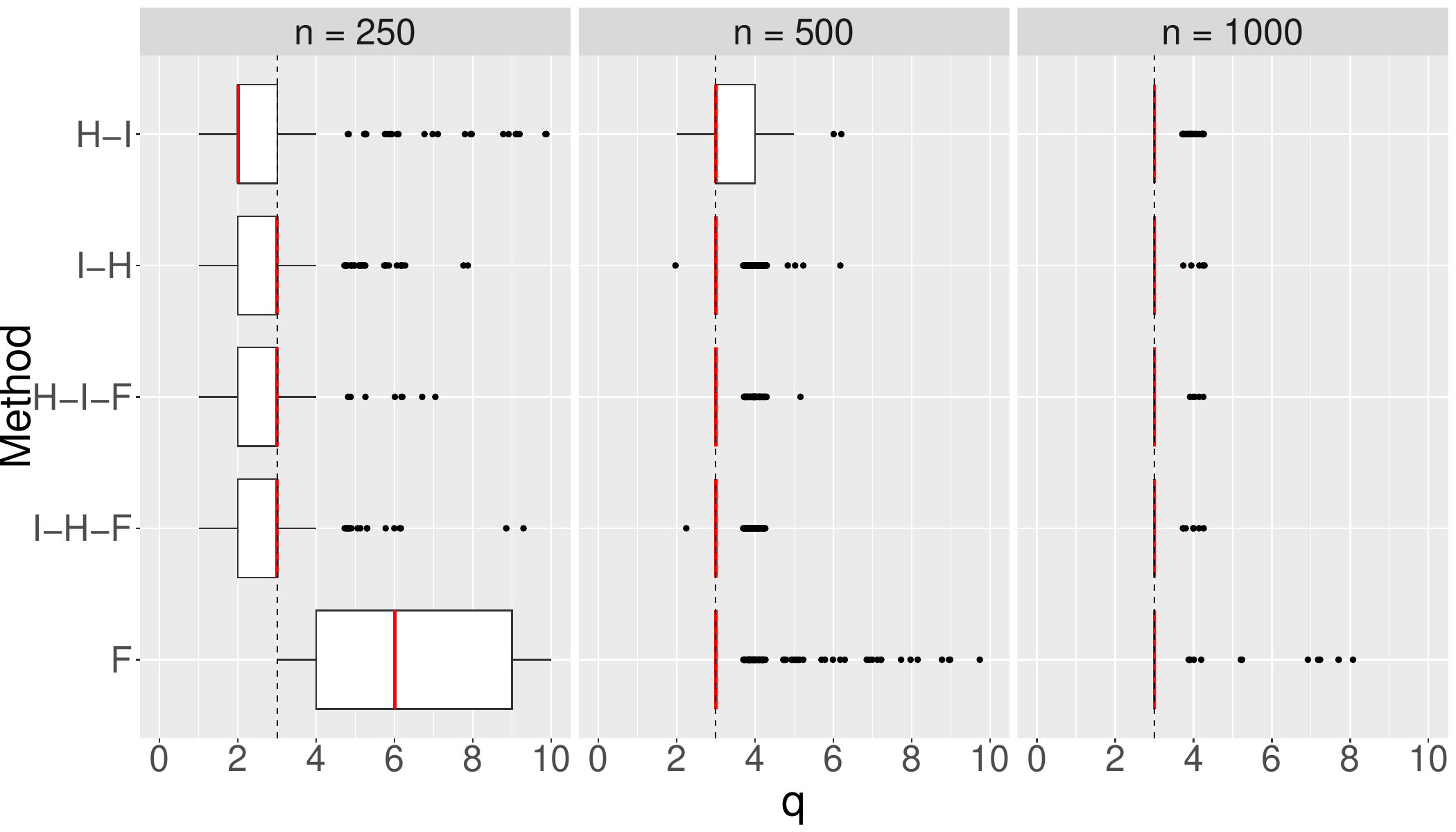}
    \caption{Simulation 1: boxplots for $q$ (the number of hidden nodes selected) for each method by sample size. Median value highlighted in red. Dashed line indicates the true value of $q$.}
    \label{fig: sim1_boxplot_q}
\end{figure}

The objective function used for model selection is BIC, and each approach has $n_\text{init} = 5$ initial vectors for the optimisation procedure.
(The choice of objective function and the effect of $n_\text{init}$ are investigated in Section \ref{sec: sim2} and Appendix \ref{app: sim4}, respectively.)
The results of the simulation study are shown in Table~\ref{tab: approaches}.
Boxplots for TNR for the inputs and $q$ for all approaches are displayed in Figure \ref{fig: sim1_boxplot_C} and Figure \ref{fig: sim1_boxplot_q}, respectively.
The true-positive rate is not shown as it is one for all methods.

\begin{table}[ht!]\centering
\begin{threeparttable}
\caption{Simulation 1: model selection metrics.}
\label{tab: approaches} 
\begin{tabular}{ccccccccc}
  \toprule
       & &           & \multicolumn{3}{c}{Input layer}  & \multicolumn{2}{c}{Hidden layer}  \\
  $n$  & Method & Time (s) & TNR & FDR & PI &  $\bar{q}$ (3) & PH &  PT \\
   \cmidrule(lr){1-1} \cmidrule(lr){2-2}  \cmidrule(lr){3-3} \cmidrule(lr){4-6} \cmidrule(lr){7-8} \cmidrule(lr){9-9}
       & H-I   & \textbf{13}  & 0.78           & 0.23                  & 0.59       & 2.29          & 0.18 & 0.10 \\
       & I-H   & 50           & 0.25           & 0.70                   & 0.01      & 2.85         & 0.44 & 0.01 \\ 
  250  & H-I-F & 14           & \textbf{0.87}  & \textbf{0.15}          & \textbf{0.72} 
       & 2.66 & \textbf{0.54} & \textbf{0.43} \\
       & I-H-F & 53           & 0.46           & 0.61                   & 0.03        & \textbf{2.87} & 0.50         & 0.03 \\
       & F     & 116         & 0.77           & 0.29                   & 0.47       & 8.58         & 0.13 & 0.12 \\
   \cmidrule(lr){1-1} \cmidrule(lr){2-2}  \cmidrule(lr){3-3} \cmidrule(lr){4-6} \cmidrule(lr){7-8} \cmidrule(lr){9-9}
       & H-I   & \textbf{32}  & 0.90           & 0.10                  & 0.83       & 3.47  & 0.53        & 0.50 \\ 
       & I-H   & 100           & 0.64           & 0.36                   & 0.42      & 3.14  & 0.87        & 0.40 \\ 
  500  & H-I-F & 36           & 0.96           & 0.05                   & \textbf{0.90}  & \textbf{3.05} & \textbf{0.95} & \textbf{0.85} \\ 
       & I-H-F & 103          & 0.72           & 0.32                   & 0.46       & 3.08  & 0.92        & 0.43 \\ 
       & F     & 82           & \textbf{0.97}  & \textbf{0.04}          & 0.89       & 3.17  & 0.90       & 0.82 \\ 
   \cmidrule(lr){1-1} \cmidrule(lr){2-2}  \cmidrule(lr){3-3} \cmidrule(lr){4-6} \cmidrule(lr){7-8} \cmidrule(lr){9-9}
       & H-I   & \textbf{53}          & \textbf{1.00}                                    & \textbf{0.00}          & \textbf{0.99}  & 3.02         & 0.98 & 0.97 \\
       & I-H   & 186          & 0.87           & 0.14          & 0.78        & \textbf{3.00}          & \textbf{1.00} & 0.77 \\
 1000  & H-I-F & \textbf{53}  & \textbf{1.00}  & \textbf{0.00}          & \textbf{0.99}  & \textbf{3.00} & \textbf{1.00} & \textbf{0.99} \\ 
       & I-H-F & 189          & 0.88                                    & 0.14          & 0.77      & 3.01 & 0.99  & 0.76 \\ 
       & F     & 169          & 0.99                                    & 0.02          & 0.97       & 3.04     & 0.99     & 0.96 \\ 
  \bottomrule
\end{tabular}
{\footnotesize\begin{tablenotes}[para, flushleft]
	\item[]{Time (s), median time to completion in seconds (carried out on an Intel\textsuperscript{\textregistered} Core\textsuperscript{\texttrademark}  i5-10210U Processor).
	Best values for a given sample size are highlighted in \textbf{bold}.}
	\end{tablenotes}}
\end{threeparttable}
\end{table}

Looking at the the model selection metrics, it is clear that the proposed H-I-F approach performs well, both in terms of selecting the correct set of input nodes and selecting the correct number of hidden nodes.
Furthermore, the TNR is high, the FDR is low, and, as expected, we see that performance improves across all metrics with increasing sample size.
From the results in Table \ref{tab: approaches}, and from Figures \ref{fig: sim1_boxplot_C} and \ref{fig: sim1_boxplot_q}, it is clear that the H-I-F approach performs best at recovering the true model structure.

Comparing the methods without the fine-tuning stage in the boxplots, and looking at layerwise selection, the probability of selecting the correct structure is increased when that layer is selected in the second phase, e.g., input-node selection is best when it comes second (see H-I versus I-H in Figure \ref{fig: sim1_boxplot_C}).
This suggests a relationship between the structure of the input and hidden layers (the probability of correctly selecting the structure of one layer increases when the other layer is more correctly specified).
This is investigated further in Appendix \ref{app: rel_structure}.
Therefore, H-I is likely better than I-H due to input-node selection being a more difficult task than hidden-node selection (determining the optimal \emph{set} of input nodes versus the optimal \emph{number} of hidden nodes), and, hence, it is favourable to perform it after hidden-node selection (given the number of hidden nodes is not substantially larger than the number of input nodes).
This relationship between the structure of both layers can be handled by incorporating a fine-tuning phase after both the H and I phases are completed.
Recall that the aim of fine tuning is to search for an improved solution in a neighbourhood of the current solution, where both H and I steps are carried out alternately (and include both backward and forward selections).
Indeed, we see that the addition of the fine-tuning phase improves on H-I in the smaller sample sizes (in large part due to improved hidden-layer selection), but its addition does not greatly improve on I-H.
Moreover, a boxplot for the computational time for each approach is provided in Appendix \ref{app: sim1_comptime}, and the addition of fine tuning only marginally adds to the computational expense.
Overall, H-I-F is significantly better than I-H-F both in terms of computational expense and model selection.
One may also consider only carrying out fine-tuning steps from the offset, which we denote by F.
However, this does not perform as well as H-I-F at the smallest sample size and is more computationally demanding.
From the above, the H-I-F approach is what we suggest as it leads to good model selection performance while also being the most computationally efficient approach.

\subsection{Simulation 2: Model Selection Objective Function}\label{sec: sim2}
This simulation study aims to determine the performance of using different objective functions when carrying out model selection.
In particular, it aims to determine whether the use of an information criterion can improve the ability for the model selection procedure to recover the true model; this is compared to the far more common approach in neural networks of using out-of-sample performance.
Three objective functions are investigated: BIC, AIC, and out-of-sample mean squared error (OOS).
The AIC approach is the same as the proposed approach in Section \ref{sec: approach}, swapping BIC for AIC $= -2\ell(\hat{\theta}) + 2(K+1)$.
The OOS approach follows the same procedure, but with the objective function replaced by out-of-sample mean squared error, which is calculated on an additional validation data set that is 20\% the size of the training data set, i.e., $\text{OOS} = \frac{1}{\Tilde{n}}\sum_{i=1}^{\Tilde{n}}(\Tilde{y}_i - \text{NN}(\Tilde{x}_i))^2$, where $\tilde n$ is the number of observations in the validation data set with response variable $\tilde y_i$ and covariate vector $\tilde x_i$.
As before, $n_\text{init} = 5$ random initialisations are used.
The results of the simulation study are shown in Table~\ref{tab: objective_function} and boxplots of TNR for the inputs and $q$ for the different objective functions are given in Appendix \ref{app: sim2_boxplots}.

\begin{table}[t]\centering
\begin{threeparttable}
\caption{Simulation 2: model selection metrics.}
\label{tab: objective_function} 
\begin{tabular}{cccccccccc}
  \toprule
       &   &          \multicolumn{3}{c}{Input layer}  & \multicolumn{2}{c}{Hidden layer}  \\
  $n$  & Method & TNR  & FDR & PI & $\bar{q}$ (3) & PH &  $K$ (16) & OOS Test &  PT \\
   \cmidrule(lr){1-1} \cmidrule(lr){2-2} \cmidrule(lr){3-5}  \cmidrule(lr){6-7} \cmidrule(lr){8-10} 
         & AIC   & 0.25 & 0.71 & 0.00 &  11.70 & 0.00 & 144 & 2.29 &  0.00   \\
  250    & BIC   & \textbf{0.87} & \textbf{0.15} & \textbf{0.72} &  2.66  & \textbf{0.54} & \textbf{16}  & \textbf{0.86} &  \textbf{0.43}   \\
         & OOS  & 0.45 & 0.60 & 0.04  & \textbf{2.79} & 0.28 & 27 & 1.30 & 0.01 \\
   \cmidrule(lr){1-1} \cmidrule(lr){2-2} \cmidrule(lr){3-5}  \cmidrule(lr){6-7} \cmidrule(lr){8-10} 
         & AIC   & 0.24 & 0.71 & 0.00 &  11.40 & 0.00 & 144 & 1.03 &  0.00   \\
  500    & BIC   & \textbf{0.96} & \textbf{0.05} & \textbf{0.90} & \textbf{3.05}  & \textbf{0.95} & \textbf{16}  & \textbf{0.53} &  \textbf{0.85}   \\
         & OOS  & 0.46 & 0.60 & 0.03 &  3.91 & 0.36 & 37 & 0.57 & 0.00\\
   \cmidrule(lr){1-1} \cmidrule(lr){2-2} \cmidrule(lr){3-5}  \cmidrule(lr){6-7} \cmidrule(lr){8-10} 
         & AIC   & 0.27 & 0.70 & 0.00 &  11.40 & 0.00 & 141 & 0.76 &  0.00   \\
  1000   & BIC   & \textbf{1.00} & \textbf{0.00} & \textbf{0.99}  & \textbf{3.00} & \textbf{1.00} & \textbf{16}  & \textbf{0.56} &  \textbf{0.99}   \\
         & OOS  & 0.53 & 0.57 & 0.02 &  3.72 & 0.46 & 34 & 0.57 & 0.00 \\
  \bottomrule
\end{tabular}
{\footnotesize\begin{tablenotes}[para, flushleft]
	\item[]{
	Best values for a given sample size are highlighted in \textbf{bold}.}
	\end{tablenotes}}
\end{threeparttable}
\end{table}

\begin{figure}[t!]
    \centering
    \includegraphics[width=14cm]{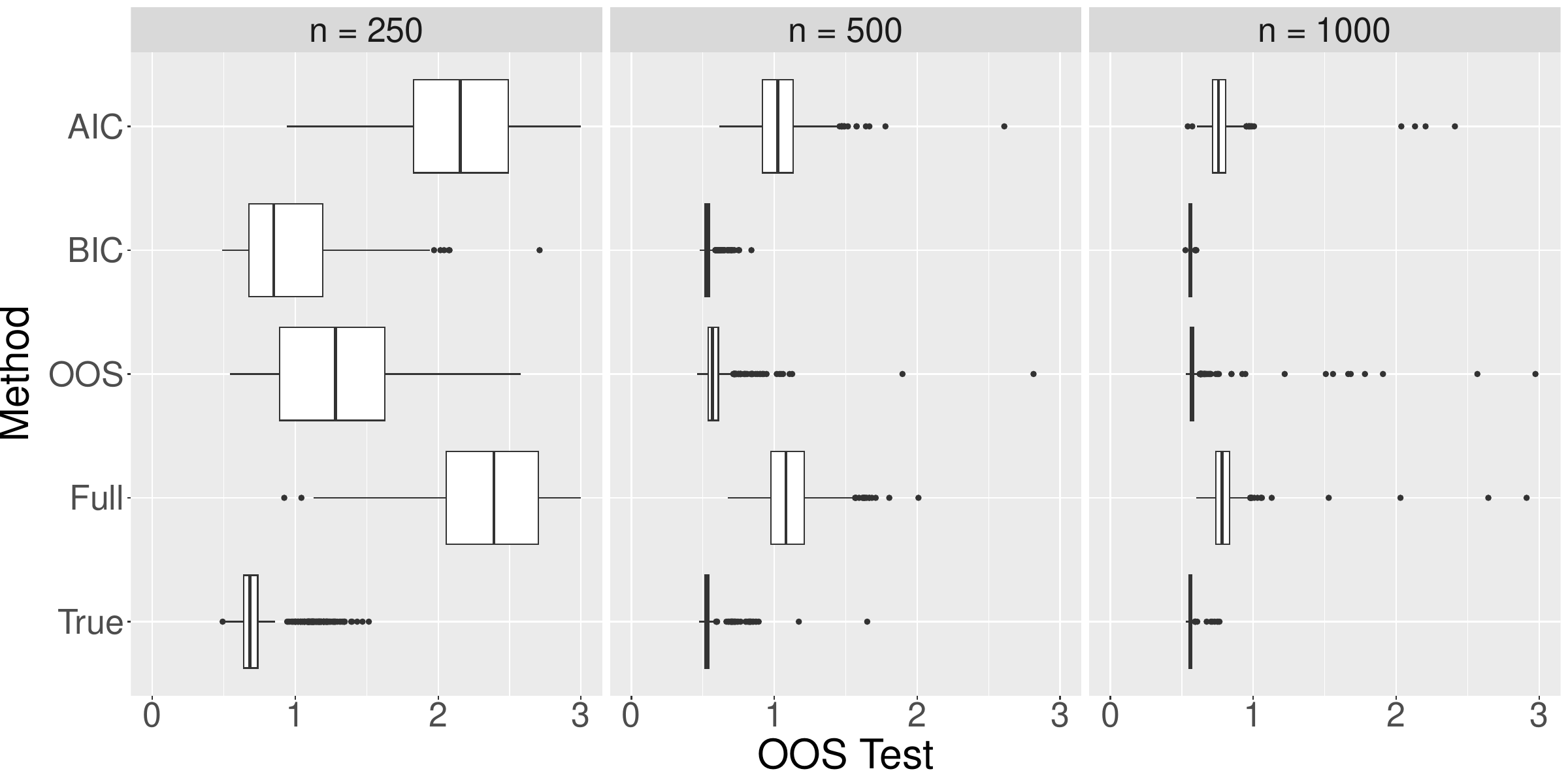}
    \caption{Simulation 2: boxplots for OOS Test for the models selected by each objective function; for comparison, the results for the true model (with inputs $x_1, x_2, x_3$ and $q = 3$) and the full model (with inputs $x_1, x_2, \dotsc, x_{13}$ and $q=10$).}
    \label{fig: sim2_boxplot_oostest}
\end{figure}

\begin{figure}[t!]
    \centering
    \includegraphics[width=14cm]{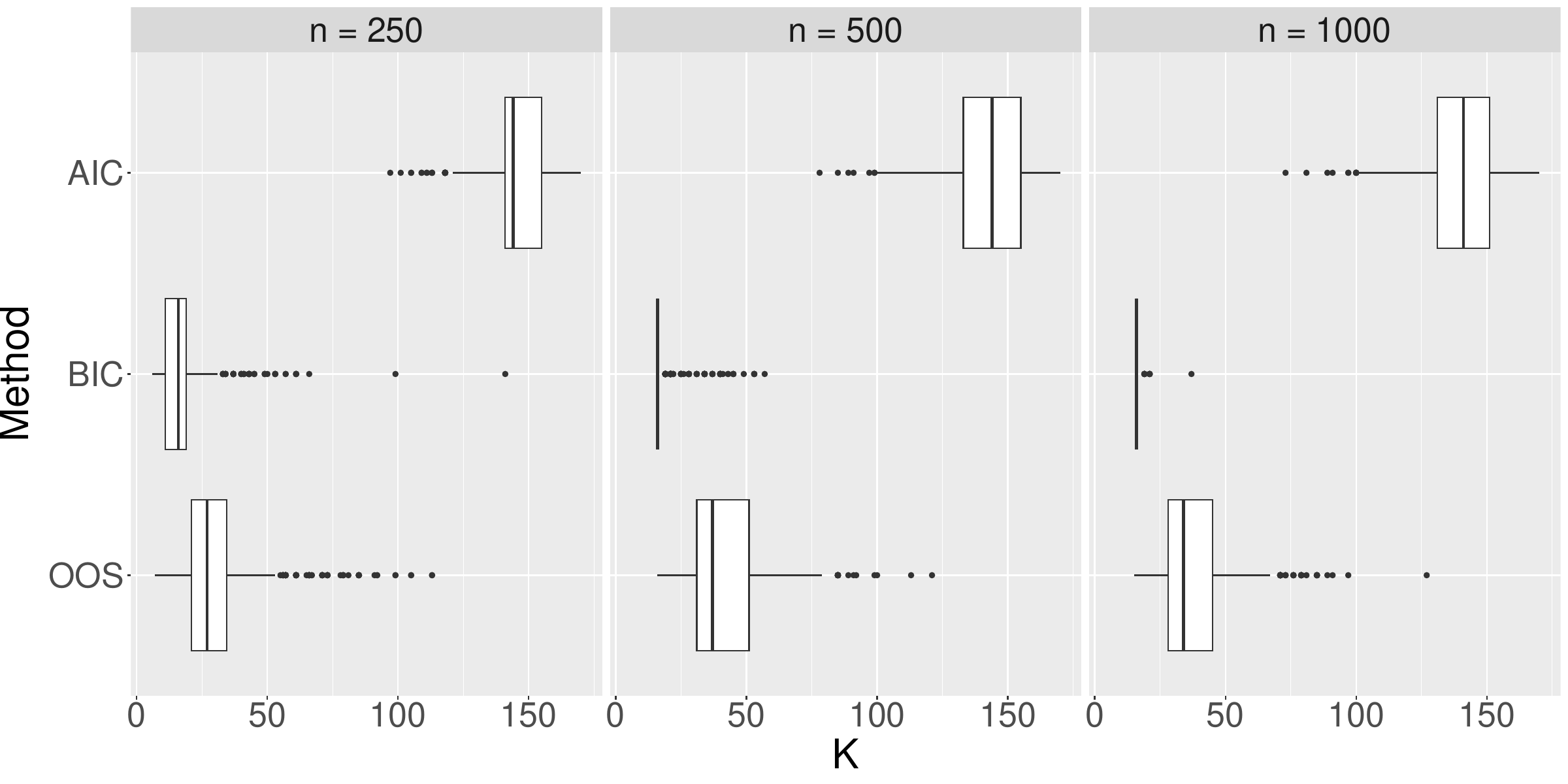}
    \caption{Simulation 2: boxplots for $K$ (number of parameters) for the models selected by each objective function.}
    \label{fig: sim2_boxplot_K}
\end{figure}

The results show that BIC far outperforms OOS and AIC in correctly identifying the correct FNN architecture.
Using OOS as the model selection objective function almost never leads to correct neural network architecture being identified.
This is due to the inability of the OOS to correctly identify and remove the unimportant covariates (TNR is always relatively low).
Using AIC leads to even worse performance, and this is likely due to the weaker penalty on model complexity compared to BIC.
It is also of interest to compare the approaches in terms of the size of the model selected and its out-of-sample performance.
The median number of neural network parameters, $K$ (note that the true value is $K = 16$), and the median out-of-sample mean squared error (OOS Test) evaluated on a test set are reported.
The OOS Test is computed on an entirely new dataset (20\% the size of the training set) that the OOS-optimising procedure was not exposed to. 
Interestingly, BIC-minimisation leads to the lowest OOS values on the test data.
This is particularly noteworthy since this is achieved using approximately half as many parameters as the OOS-minimisation procedure.
Boxplots highlighting the values of OOS Test and $K$ are shown in Figures \ref{fig: sim2_boxplot_oostest} and \ref{fig: sim2_boxplot_K}, respectively.
Figure \ref{fig: sim2_boxplot_oostest} also displays the OOS Test values for the true model (inputs $x_1, x_2, x_3$ and $q=3$) and the full model (inputs $x_1, x_2, \dotsc, x_{13}$ and $q=10$); this allow us to evaluate the performance of selection compared to the full model, and how close we can get to the true model.
The models selected using the BIC procedure have similar performance to the true model, particularly as the sample size increases.
In contrast, the models selected using AIC have worse out-of-sample performance and significantly more parameters, and the performance is similar to fitting the full model.

We have also compared our proposed BIC-based selection procedure to two commonly used strategies for dealing with overfitting, namely, weight decay and early stopping. The results are deferred to Appendix \ref{app: wd_and_es}, where we have found that our proposed approach yields improved OOS Test values compared to these other two strategies.

\subsection{Simulation 3: Data-generating Process is not a Neural Network}\label{sec: sim3}
For this simulation study, we investigate the performance of the proposed H-I-F model selection procedure on a data set simulated from a data-generating process that is not a neural network:
\begin{equation}\label{eq: data_gen}
    y = x_1 - 0.75 x_2^2 + 0.9 x_3x_4x_5 + \varepsilon,
\end{equation}
where $x_1, x_2, \dotsc, x_{10} \sim N(0, 1)$, i.e., there are five relevant and five irrelevant covariates, and $\sigma^2 = Var(\varepsilon) = 0.3$.
For comparison, we have also performed stepwise model selection for a linear model using BIC.
We applied this using the \texttt{stepAIC} function from the \texttt{MASS} R package with \texttt{k = log(n)} \citep{masspackage}.
To compare with the H-I-F procedure, we also performed stepwise selection on a linear model with a search space containing (i) all terms up to three-way interactions (step-lm-3), (ii) all terms up to two-way interactions (step-lm-2), and (iii) only main effects (step-lm-1). Note that the first model is correctly specified, and the latter two are misspecified.
For these linear models, we began the search with all possible terms in the model, and allowed the stepwise search to consider both the elimination of an included variable and the addition of an excluded variable at each step (i.e., \texttt{direction = "both"}).
For the purpose of this study, when computing performance metrics (displayed in Table \ref{tab: corr1}), we only considered whether or not relevant variables ($x_1, \ldots, x_5$) and irrelevant variables ($x_6, \ldots, x_{10}$) are selected. While the exact functional form of each selected variable is not considered, the OOS metrics facilitate model comparisons in the sense that lower OOS values imply a better approximation to the generating model (i.e., the functional form of input variables).
In Table \ref{tab: corr1}, as with earlier tables, the TNR, FDR, and PT selection metrics are shown, but, here, the TPR (true positive rate) metric is also shown.
Moreover, we also show median number of parameters ($K$), the median out-of-sample mean squared error evaluated on a test set (OOS Test), and the median computational time (Time) for each approach.

\begin{table}[th!]\centering
\begin{threeparttable}
\caption{Simulation 3: Comparison of proposed model selection approach for neural networks with stepwise model selection for linear models for the data-generating process given by Equation \ref{eq: data_gen}.}
\label{tab: corr1} 
\begin{tabular}{ccccccccc}
  \toprule
  Method & $n$  & TPR & TNR & FDR & PT & $K$ & OOS Test & Time (s) \\
  \cmidrule(lr){1-1}  \cmidrule(lr){2-2} \cmidrule(lr){3-6}  \cmidrule(lr){7-9} 
H-I-F & 250  & 0.53 & 0.90 & 0.13 & 0.02 & 11  & 2.12 & 12 \\
H-I-F & 500  & 0.78 & 0.95 & 0.04 & 0.50 & 43 & 0.52 & 22 \\
H-I-F & 1000 & 1.00 & 0.98 & 0.02 & 0.93 & 57 & 0.30 & 46\\
  \cmidrule(lr){1-1}  \cmidrule(lr){2-2} \cmidrule(lr){3-6}  \cmidrule(lr){7-9} 
step-lm-3 & 250  & 1.00 & 0.13 & 0.45 & 0.04 & 61  & 0.73 & 53 \\
step-lm-3 & 500  & 1.00 & 0.46 & 0.32 & 0.12 & 20  & 0.36 & 105 \\
step-lm-3 & 1000 & 1.00 & 0.63 & 0.24 & 0.23 & 16 & 0.28 & 215 \\
  \cmidrule(lr){1-1}  \cmidrule(lr){2-2} \cmidrule(lr){3-6}  \cmidrule(lr){7-9} 
step-lm-2 & 250  & 0.85 & 0.35 & 0.43 & 0.00 & 15  & 1.95 & 2 \\
step-lm-2 & 500  & 0.83 & 0.39 & 0.42 & 0.00 & 13 & 1.31 & 3 \\
step-lm-2 & 1000 & 0.93 & 0.80 & 0.16 & 0.00 & 10 & 1.02 & 5\\
  \cmidrule(lr){1-1}  \cmidrule(lr){2-2} \cmidrule(lr){3-6}  \cmidrule(lr){7-9} 
step-lm-1 & 250  & 0.20 & 1.00 & 0.00 & 0.00 & 2  & 3.55 & 0\\
step-lm-1 & 500  & 0.20 & 1.00 & 0.00 & 0.00 & 2 & 2.46 & 0\\
step-lm-1 & 1000 & 0.40 & 1.00 & 0.00 & 0.00 & 3 & 1.98 & 0 \\
  \bottomrule
\end{tabular}
{\footnotesize\begin{tablenotes}[para, flushleft]
	\item[]{Time (s), median time to completion in seconds (carried out on an Intel\textsuperscript{\textregistered} Core\textsuperscript{\texttrademark}  i5-10210U Processor).}
	\end{tablenotes}}
\end{threeparttable}
\end{table}

From Table \ref{tab: corr1}, we see that the proposed H-I-F procedure has a high true negative rate, a low false discovery rate, and the true positive rate increases with the sample size; consequently, the probability of selecting the true set of covariates (PT) increases with the sample size. At the highest sample size, the out-of-sample performance is very close to that of the correctly specified third order linear model (step-lm-3).
Although this true step-lm-3 model provides the lowest out-of-sample performance, its true negative and false discovery rates are relatively poor compared to the neural network, and, hence, the probability of selecting the true set of covariates does not approach one for the sample sizes we have considered. The selected step-lm-3 model does have fewer parameters on average than the neural network model (at $n = 500$ and $n = 1000$), but the step-lm-3 search is far more computationally intensive; this is due to the large number of possible interaction terms up to order three. 
It is important to note that the stepwise approaches for the linear models require the search space of models to be explicitly specified through the interaction and polynomial terms, and the performance of the misspecified (step-lm-2 and step-lm-1) approaches is very poor.
In contrast, the proposed H-I-F selection approach does not require these terms to be explicitly specified, but still achieves very good out-of-sample performance since complex functional relationships and interactions are captured in a more automatic manner within the neural network structure.

\section{Application to Data}\label{sec: app_to_data}
Airbnb is an online marketplace that provides both short-term and long-term rentals.
Data relating to the rental listings can be obtained from Inside Airbnb (\url{http://insideairbnb.com}).
Here, we focus on rental listings in the Dún Laoghaire--Rathdown area of Dublin on the seventh of September 2023, and aim to implement our proposed model selection approach, and determine factors that may be associated with the listing's price.
The data consists of information relating to 625 rental listings, and the following explanatory variables: the number of people the rental accommodates (\texttt{accommodates}), the rental's review rating (\texttt{rating}), the number of reviews per month (\texttt{num\_reviews}), an indicator of whether the rental is an entire home or a private room (\texttt{room\_type}; 0 for an entire home, 1 for a private room), an indicator of whether or not the host is a ``superhost" (\texttt{superhost}; i.e., top-performing Airbnb hosts, where performance is based on reviews, responsiveness and their cancellation rate), the total number of Airbnb listings that the host has (\texttt{num\_listings}), an indicator of whether or not the listing is instantly bookable (\texttt{instant}), and the latitude (\texttt{latitude}) and longitude (\texttt{longitude}) of the rental.
The response variable is the natural logarithm of the price per night of each rental (\texttt{lnprice}).
The data is available in our R package \texttt{selectnn} \citep{mcinerney2022selectnn}.

The dataset has been randomly split into a training set and test set with a 80\%--20\% split, respectively, and all continuous variables have been standardised (based on the training data) to have zero mean and unit variance.
The model selection procedure was implemented with $n_{\text{init}} = 10$ and $q_{\text{max}} = 10$.
For comparison purposes, the model found by our proposed model selection procedure is compared to fitting an FNN with all inputs and the maximum number of hidden nodes considered.
For both models (selected and full), we report the number of input nodes ($p$), the number of hidden nodes ($q$), the total number of parameters ($K$), the BIC, and the out-of-sample mean squared error (OOS) computed using the test set.
For the covariates that are selected, we also report: (i) relative covariate importance via the change in BIC ($\Delta$BIC) upon removal of that covariate, and (ii) a simple covariate effect ($\hat \tau$) as measured by the change in the average predicted response going from lower to higher covariate values (below/above median for numeric covariates and 0/1 for binary covariates).
See Appendix \ref{app: delta_bic} for more detail on these measures.

Our proposed procedure selects two hidden nodes and includes three covariates: \newline \texttt{accommodates}, \texttt{num\_reviews} and \texttt{room\_type}.
As shown in Table~\ref{tab: airbnb_sel}, the selected model has 100 fewer parameters than the full model, while also having a much lower BIC value and a lower out-of-sample mean squared error.
The BIC differences and covariate effects (and their associated bootstrapped confidence intervals) for the variables that remain in the model are reported in Table \ref{tab: airbnb_eff}.
Using $\Delta\text{BIC}$ as a measure of variable importance, we find that \texttt{accommodates} is the most important variable with $\Delta\text{BIC}_{\texttt{rm}} = 200.16$.
Based on its effect ($\hat{\tau}_{\texttt{accommodates}} = 1.38$), the more people the listing accommodates, the higher the price per night.
The binary variable \texttt{room\_type} has a negative effect, which suggests that the listing price of a private room is lower than an entire house, on average.
The other covariate, \texttt{num\_reviews}, is more important than \texttt{room\_type} as judged by its $\Delta$BIC value, but the confidence interval for its covariate effect includes zero. This suggests that \texttt{num\_reviews} has a non-linear effect that cannot be seen in an overall average change in the predicted response.

\begin{table}
    \centering
    \caption{Dublin Airbnb: selected versus full model comparison.}
   \begin{tabular}[c]{lccccc}
  \toprule
   & $p$ & $q$ & $K$ & BIC & OOS  \\
   \cmidrule(lr){2-6} 
  Selected & 3 & 2 & 11 & 884.2 & 0.25 \\
  Full & 9 & 10 & 111 & 1136.3 & 0.48 \\
  \bottomrule
\end{tabular}\label{tab: airbnb_sel}
  \end{table}

\begin{table}[t!]
    \centering
    \caption{Dublin Airbnb: covariate effects and BIC differences.}
    \begin{tabular}[c]{lcc}
      \toprule
     & $\hat{\tau}$ (95\% CI) & $\Delta\text{BIC}$  \\ 
     \cmidrule(lr){2-2} \cmidrule(lr){3-3}
  \texttt{accommodates} & ~1.38 (~1.30, ~1.47)  & 200.16 \\
  \texttt{num\_reviews} & ~0.07 (-0.07, ~0.21)  & 89.56 \\
  \texttt{room\_type} & -1.44 (-1.51, -1.37) & 64.02 \\
       \bottomrule
    \end{tabular}\label{tab: airbnb_eff}
  \end{table}

The selected model dropped six covariates (from a set of nine possible covariates).
While the underlying selection procedure cannot guarantee that this model minimises the BIC, and an exhaustive search through all sub-models is computationally expensive, we have nevertheless carried out such a search for the purpose of comparison.
To this end, we fitted the model with all covariates included, all nine models that arise by dropping one of each of the covariates, all 36 models that arise when pairs of covariates are dropped, all 84 models that arise when triples of covariates are dropped, and so on, for each hidden layer size, $q = 1, \dotsc, 10$.
Each model was allowed $n_{\text{init}} = 10$ random initialisations, mirroring that of our selection procedure.

\begin{figure}[t]
\centering
\includegraphics[width=\textwidth]{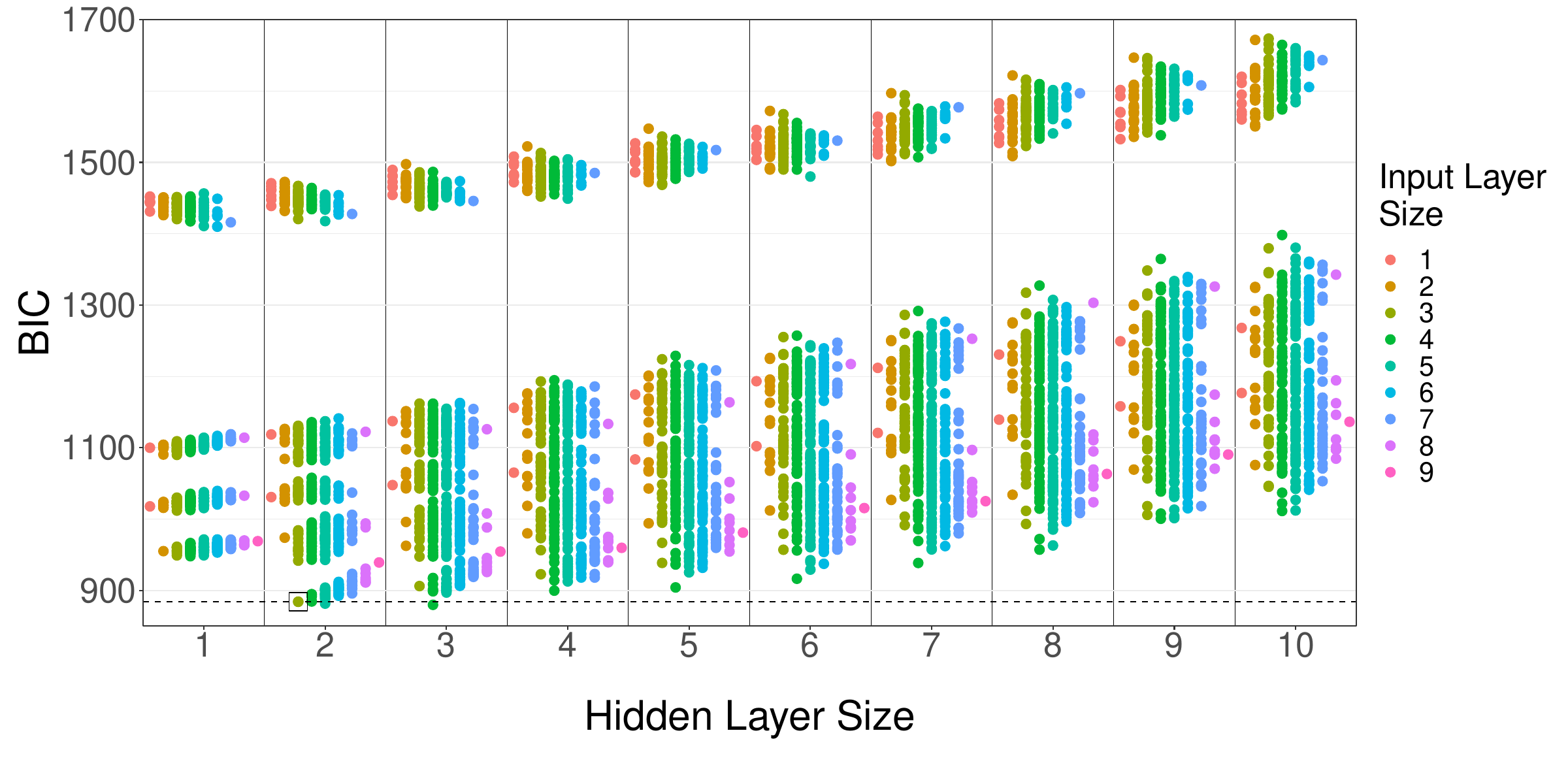}
\caption{Dublin Airbnb: BIC of models for different input-layer and hidden-layer combinations. Points are coloured according to the input layer size. The model selected by our procedure is enclosed in a box and the horizontal dashed line indicates the BIC for this model.}
\label{fig: bic_airbnb}
\end{figure}

Figure~\ref{fig: bic_airbnb} shows the BIC for each model where each point corresponds to a different input-layer hidden-layer combination; for comparison, the model selected by our procedure is indicated using a box.
First, note that there is a subset of models with relatively large BIC values.
Each of these models are missing the variables \texttt{accommodates} and \texttt{room\_type}, which further highlights their importance in the model.
It is clear  that the proposed selection procedure has indeed found a model with a BIC value that is among the lowest of the alternative models we have considered.
That being said, the exhaustive search did return two models with lower BIC values ($\Delta\text{BIC} = -4.4$ and $\Delta\text{BIC} = -2.7$).
These models are more complex than the model selected by our procedure (with $q = 2$ and $p = 3$) as they have $q = 3$ and $q = 2$ hidden nodes with $p = 4$ and $p = 5$ input nodes, respectively.
We note that the out-of-sample predictive performance is very similar across all three of these models.

\section{Discussion}\label{sec: disc}
FNNs have become very popular in recent years and have the potential to capture more complex covariate effects than traditional statistical models.
However, model selection procedures are of the utmost importance in the context of FNNs since their flexibility may increase the chance of over-fitting; indeed, the principle of parsimony is very common throughout statistical modelling more generally. 
Therefore, we have proposed a statistically-motivated neural network selection procedure by assuming an underlying (normal) error distribution, which then permits BIC minimisation.
More specifically, our procedure involves a hidden-node selection phase, followed by an input-node (covariate) selection phase, followed by a final fine-tuning phase.
We have made this procedure available in our \texttt{selectnn} package in R \citep{mcinerney2022selectnn}.

Through extensive simulation studies, we have found that that (i) the order of selection (input versus hidden layer) is important, with respect to the probability of recovering the true model and the computational efficiency, (ii) the addition of a fine-tuning stage provides a non-negligible improvement while not significantly increasing the computational burden, (iii) using the BIC is necessary to asymptotically converge to the true model, and (iv) although the models selected using BIC have fewer parameters than those selected using out-of-sample performance, they have comparable, and sometimes improved, predictivity.
We suggest that statistically-orientated model selection approaches are necessary in the application of neural networks --- just as they are in the application of more traditional statistical models --- and we have demonstrated the favourable performance of our proposal.

In its current form, a limitation of the proposed procedure is that, due to its stepwise nature, it would be more computationally intensive when dealing with larger models and datasets.
We expect that randomisation and/or divide-and-conquer throughout the selection phases would be required in more complex problems involving many covariates and/or hidden layers, and adaptations may also be required for stochastic optimisation procedures used on much larger datasets.
Nevertheless, neural networks are still valuable in more traditional (smaller) statistical problems for which procedures such as ours will lead to more insightful outputs.
Furthermore, the implementation of statistical approaches more broadly (such as uncertainty quantification and hypothesis testing) in neural network modelling will be crucial for the enhancement of these insights.
This will be the direction of our future work.

\section*{Acknowledgement}
This publication has emanated from research conducted with the financial support of Science Foundation Ireland under Grant number 18/CRT/6049. 
The second author was supported by the Confirm Smart Manufacturing Centre (https://confirm.ie/) funded by Science Foundation Ireland (Grant Number: 16/RC/3918).
For the purpose of Open Access, the author has applied a CC BY public copyright licence to any Author Accepted Manuscript version arising from this submission.

\bibliographystyle{apalike}
\bibliography{main}

\begin{thebibliography}{}

\bibitem[Abiodun et~al., 2018]{state_of_nn2018}
Abiodun, O.~I., Jantan, A., Omolara, A.~E., Dada, K.~V., Mohamed, N.~A., and Arshad, H. (2018).
\newblock State-of-the-art in artificial neural network applications: A survey.
\newblock {\em Heliyon}, 4(11):e00938.

\bibitem[Akaike, 1998]{Akaike1998}
Akaike, H. (1998).
\newblock {\em Information Theory and an Extension of the Maximum Likelihood Principle}, pages 199--213.
\newblock Springer New York, New York, NY.

\bibitem[Anderson and Burnham, 2004]{anderson2004model}
Anderson, D. and Burnham, K. (2004).
\newblock Model selection and multi-model inference.
\newblock {\em Second. NY: Springer-Verlag}.

\bibitem[Bishop et~al., 1995]{bishop1995neural_9}
Bishop, C.~M. et~al. (1995).
\newblock {\em Neural networks for pattern recognition}, chapter~9, pages 353--354.
\newblock Oxford university press.

\bibitem[Breiman, 2001]{breiman2001statistical}
Breiman, L. (2001).
\newblock {Statistical Modeling: The Two Cultures (with comments and a rejoinder by the author)}.
\newblock {\em Statistical Science}, 16(3):199 -- 231.

\bibitem[Chandrashekar and Sahin, 2014]{chandrashekar2014survey}
Chandrashekar, G. and Sahin, F. (2014).
\newblock A survey on feature selection methods.
\newblock {\em Computers \& Electrical Engineering}, 40(1):16--28.

\bibitem[Cheng and Titterington, 1994]{cheng1994neural}
Cheng, B. and Titterington, D.~M. (1994).
\newblock Neural networks: A review from a statistical perspective.
\newblock {\em Statistical Science}, 9(1):2--30.

\bibitem[Cybenko, 1989]{cybenko1989approximation}
Cybenko, G. (1989).
\newblock Approximation by superpositions of a sigmoidal function.
\newblock {\em Mathematics of Control, Signals and Systems}, 2(4):303--314.

\bibitem[Efron, 2020]{efron2020prediction}
Efron, B. (2020).
\newblock Prediction, estimation, and attribution.
\newblock {\em International Statistical Review}, 88(S1):S28--S59.

\bibitem[Elder, 2003]{elder2003paradox}
Elder, J.~F. (2003).
\newblock The generalization paradox of ensembles.
\newblock {\em Journal of Computational and Graphical Statistics}, 12(4):853--864.

\bibitem[Fan and Lv, 2010]{fan2010selective}
Fan, J. and Lv, J. (2010).
\newblock A selective overview of variable selection in high dimensional feature space.
\newblock {\em Statistica Sinica}, 20(1):101--148.

\bibitem[Fisher and Russell, 1922]{fisher1922mathematical}
Fisher, R.~A. and Russell, E.~J. (1922).
\newblock On the mathematical foundations of theoretical statistics.
\newblock {\em Philosophical Transactions of the Royal Society of London. Series A, Containing Papers of a Mathematical or Physical Character}, 222(594-604):309--368.

\bibitem[Goldberg, 2016]{goldberg2016primer}
Goldberg, Y. (2016).
\newblock A primer on neural network models for natural language processing.
\newblock {\em Journal of Artificial Intelligence Research}, 57:345--420.

\bibitem[Heinze et~al., 2018]{heinze2018variable}
Heinze, G., Wallisch, C., and Dunkler, D. (2018).
\newblock Variable selection – a review and recommendations for the practicing statistician.
\newblock {\em Biometrical Journal}, 60(3):431--449.

\bibitem[Hooker and Mentch, 2021]{hooker2021bridging}
Hooker, G. and Mentch, L. (2021).
\newblock Bridging breiman's brook: From algorithmic modeling to statistical learning.
\newblock {\em Observational Studies}, 7(1):107--125.

\bibitem[Hornik et~al., 1989]{hornik1989multilayer}
Hornik, K., Stinchcombe, M., and White, H. (1989).
\newblock Multilayer feedforward networks are universal approximators.
\newblock {\em Neural Networks}, 2(5):359--366.

\bibitem[Koenen and Wright, 2024]{koenen2024interpreting}
Koenen, N. and Wright, M.~N. (2024).
\newblock Interpreting deep neural networks with the package innsight.
\newblock {\em arXiv preprint arXiv:2306.10822}.

\bibitem[LeCun et~al., 2015]{lecun2015deep}
LeCun, Y., Bengio, Y., and Hinton, G. (2015).
\newblock Deep learning.
\newblock {\em Nature}, 521(7553):436--444.

\bibitem[McInerney and Burke, 2022]{mcinerney2022selectnn}
McInerney, A. and Burke, K. (2022).
\newblock {\em selectnn: A Statistically-Based Approach to Neural Network Model Selection}.
\newblock R package version 0.0.0.9000.

\bibitem[Miller, 2002]{miller2002subset}
Miller, A. (2002).
\newblock {\em Subset selection in regression}.
\newblock chapman and hall/CRC.

\bibitem[Murata et~al., 1994]{murata1994nic}
Murata, N., Yoshizawa, S., and Amari, S. (1994).
\newblock Network information criterion-determining the number of hidden units for an artificial neural network model.
\newblock {\em IEEE Transactions on Neural Networks}, 5(6):865--872.

\bibitem[Pang et~al., 2021]{pang2021deep}
Pang, G., Shen, C., Cao, L., and Hengel, A. V.~D. (2021).
\newblock Deep learning for anomaly detection: A review.
\newblock {\em ACM Computing Surveys (CSUR)}, 54(2):1--38.

\bibitem[Pontes et~al., 2016]{pontes2016design}
Pontes, F., Amorim, G., Balestrassi, P., Paiva, A., and Ferreira, J. (2016).
\newblock Design of experiments and focused grid search for neural network parameter optimization.
\newblock {\em Neurocomputing}, 186:22--34.

\bibitem[Ripley and Venables, 2022]{ripley2022nnet}
Ripley, B. and Venables, W. (2022).
\newblock nnet: Feed-forward neural networks and multinomial log-linear models.
\newblock {\em R package version}, 7.3-17.

\bibitem[Ripley, 1993]{ripley1993statistical}
Ripley, B.~D. (1993).
\newblock Statistical aspects of neural networks.
\newblock In Nielsen, B. O.~E., Jensen, J.~L., and Kendall, W.~S., editors, {\em Networks and Chaos: Statistical and Probabilistic Aspects}, pages 40--123. Chapman \& Hall.

\bibitem[Ripley, 1994]{ripley1994neural}
Ripley, B.~D. (1994).
\newblock Neural networks and related methods for classification.
\newblock {\em Journal of the Royal Statistical Society: Series B (Methodological)}, 56(3):409--437.

\bibitem[Rügamer et~al., 2020]{rugamer2020semistructured}
Rügamer, D., Kolb, C., and Klein, N. (2020).
\newblock Semi-structured deep distributional regression: Combining structured additive models and deep learning.
\newblock {\em arXiv preprint arXiv:2002.05777}.

\bibitem[Schwarz, 1978]{schwarz1978estimating}
Schwarz, G. (1978).
\newblock Estimating the dimension of a model.
\newblock {\em The Annals of Statistics}, 6(2):461--464.

\bibitem[Sun et~al., 2022]{SUN2022109246}
Sun, Y., Song, Q., and Liang, F. (2022).
\newblock Learning sparse deep neural networks with a spike-and-slab prior.
\newblock {\em Statistics \& Probability Letters}, 180:109246.

\bibitem[Tibshirani, 1996]{tibshirani1996regression}
Tibshirani, R. (1996).
\newblock Regression shrinkage and selection via the lasso.
\newblock {\em Journal of the Royal Statistical Society: Series B (Methodological)}, 58(1):267--288.

\bibitem[Tran et~al., 2020]{tran2020bayesian}
Tran, M.-N., Nguyen, N., Nott, D., and Kohn, R. (2020).
\newblock Bayesian deep net glm and glmm.
\newblock {\em Journal of Computational and Graphical Statistics}, 29(1):97--113.

\bibitem[Venables and Ripley, 2002]{masspackage}
Venables, W.~N. and Ripley, B.~D. (2002).
\newblock {\em Modern Applied Statistics with S}.
\newblock Springer, New York, fourth edition.
\newblock ISBN 0-387-95457-0.

\bibitem[Voulodimos et~al., 2018]{voulodimos2018deep}
Voulodimos, A., Doulamis, N., Doulamis, A., and Protopapadakis, E. (2018).
\newblock Deep learning for computer vision: A brief review.
\newblock {\em Computational Intelligence and Neuroscience}, 2018.

\bibitem[White, 1989]{white1989learning}
White, H. (1989).
\newblock {Learning in Artificial Neural Networks: A Statistical Perspective}.
\newblock {\em Neural Computation}, 1(4):425--464.

\bibitem[Ye, 1998]{ye1998measuring}
Ye, J. (1998).
\newblock On measuring and correcting the effects of data mining and model selection.
\newblock {\em Journal of the American Statistical Association}, 93(441):120--131.

\end{thebibliography}

\newpage
\appendix


\section{Neural Network Degrees of Freedom}\label{app: df}

The use of the BIC for model selection introduces the question of degrees of freedom for neural networks.
In our procedure, we define the degrees of freedom to be the number of parameters in the model, $K$.
From our simulation results in Section 4 of the main paper, we see that this leads to consistent model selection.
However, other approaches to defining the degrees of freedom for neural networks exist.
\citet{ye1998measuring} defined the concept of generalised degrees of freedom (GDF) as 
\begin{equation*}
    \text{GDF} = \frac{\sum_{i=1}^n cov(y_i, \hat y_i)}{\sigma^2},
\end{equation*}
where $\hat y_i$ are the predicted values from the model; note that the computation of GDF in practice is based refitting the model many times to datasets with slightly perturbed values of $y_i$ \citep{elder2003paradox, ye1998measuring}.
\citet{murata1994nic} introduced a network information criterion (NIC) whose penalty for model complexity is given by 
\begin{equation*}
    \text{EDF} = \text{tr}(GQ^{-1}),
\end{equation*}
where $Q = E[\nabla_\theta \nabla_\theta^T \ell(\theta)]$, $G = Var[\nabla_\theta \ell(\theta)]$ and $\text{tr}(\cdot)$ denotes the trace operator.
In this Appendix, we consider the behaviour of these degrees of freedom formulae using a variety of simulations.
The first simulation study investigates the values of EDF and GDF for different neural network architectures when the model is correctly specified.
The second simulation study also investigates EDF and GDF, but for neural networks that are incorrectly specified.
In particular, we focus on cases when the neural network fit to the data is larger than the true data-generating model.
Finally, we implement both methods as the degrees-of-freedom term within our model selection procedure, and compare the results to our proposed use of the number of parameters as the degrees of freedom (i.e., the classical BIC penalty).

\begin{figure}
    \centering
    \includegraphics[width=\textwidth]{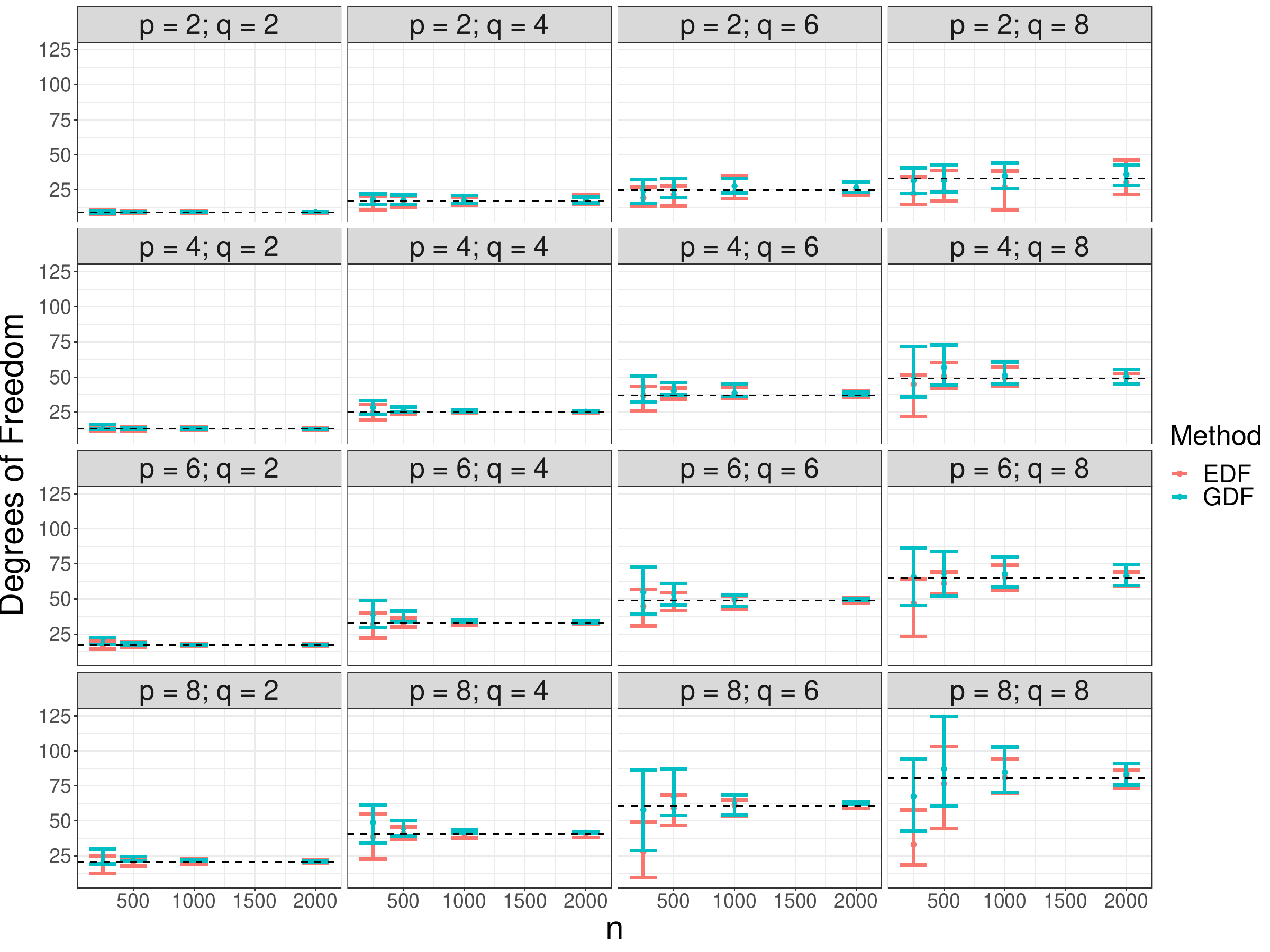}
    \caption{The average EDF and GDF values and their associated 2.5 and 97.5 quantiles versus sample size for different neural network architectures. The horizontal dashed line represents the number of parameters for each architecture.}
    \label{fig: sim_df_1}
\end{figure}

For the first simulation study, the degrees of freedom are estimated for various neural network architectures.
Both the size of the input layer and the hidden layer are varied with $p, q \in \{2, 4, 6, 8\}$.
For each architecture, a correctly specified neural network model is fit to the data.
Sample size is varied with $n \in \{250, 500, 1000, 2000\}$ and 100 simulation replicates are carried out.
The results for each architecture are displayed in Figure \ref{fig: sim_df_1}.

It is clear that the GDF, EDF, and $K$ (the number of parameters) closely align with each other in the scenarios we have considered.
In the more complicated (larger) models, there is more variability in the GDF and EDF values at smaller samples, but they converge to $K$ with the sample size. Interestingly, the convergence to $K$ is from below, implying that, if GDF or EDF are used within a BIC selection procedure, they will penalise complexity less than when using $K$ (and, hence, select more complex models than those selected using the classical BIC).

\begin{figure}
    \centering
    \includegraphics[width=13cm]{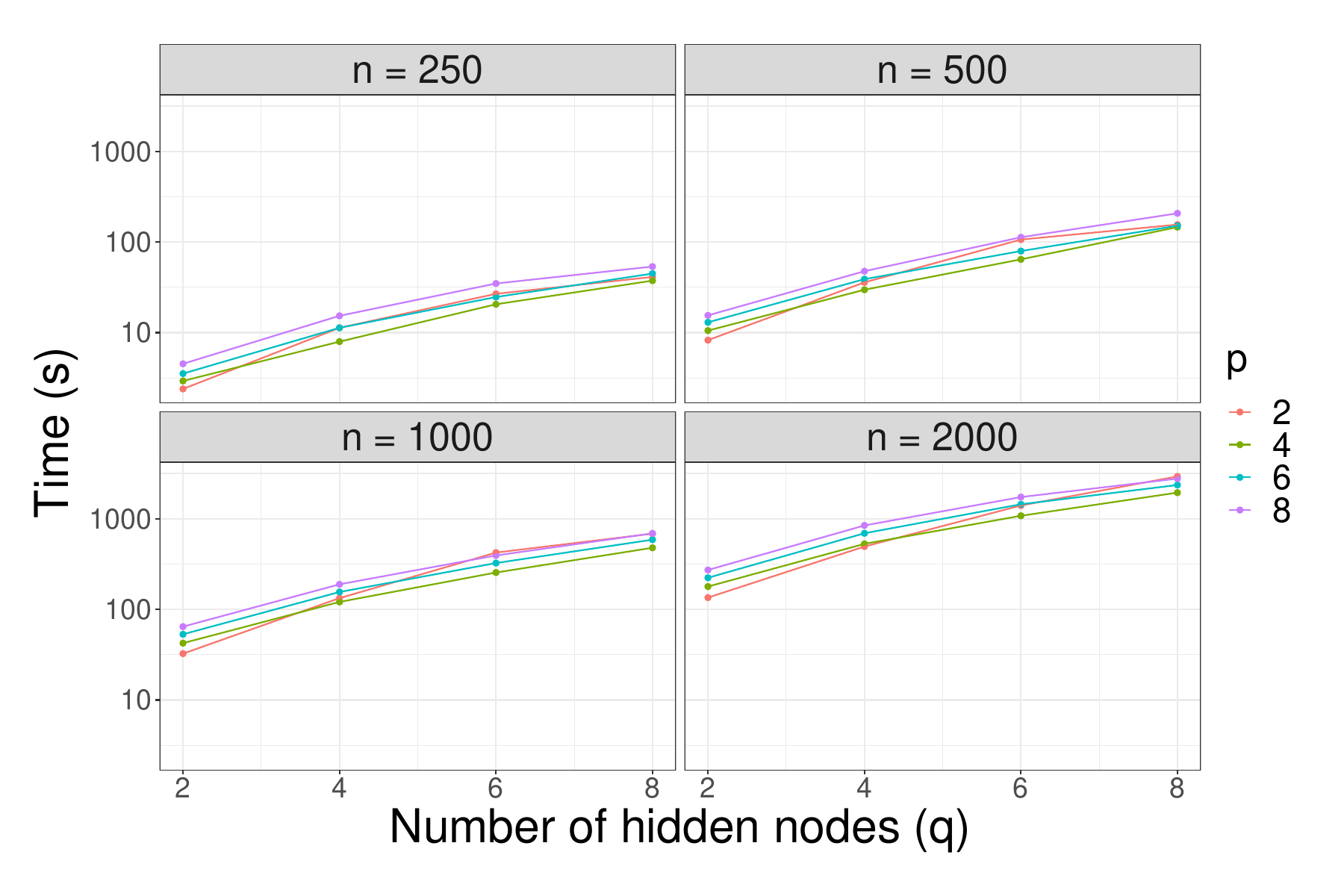}
    \caption{The average time taken to compute GDF for different architectures and sample sizes.}
    \label{fig: sim_gdf_time}
\end{figure}

\begin{figure}
    \centering
    \includegraphics[width=13cm]{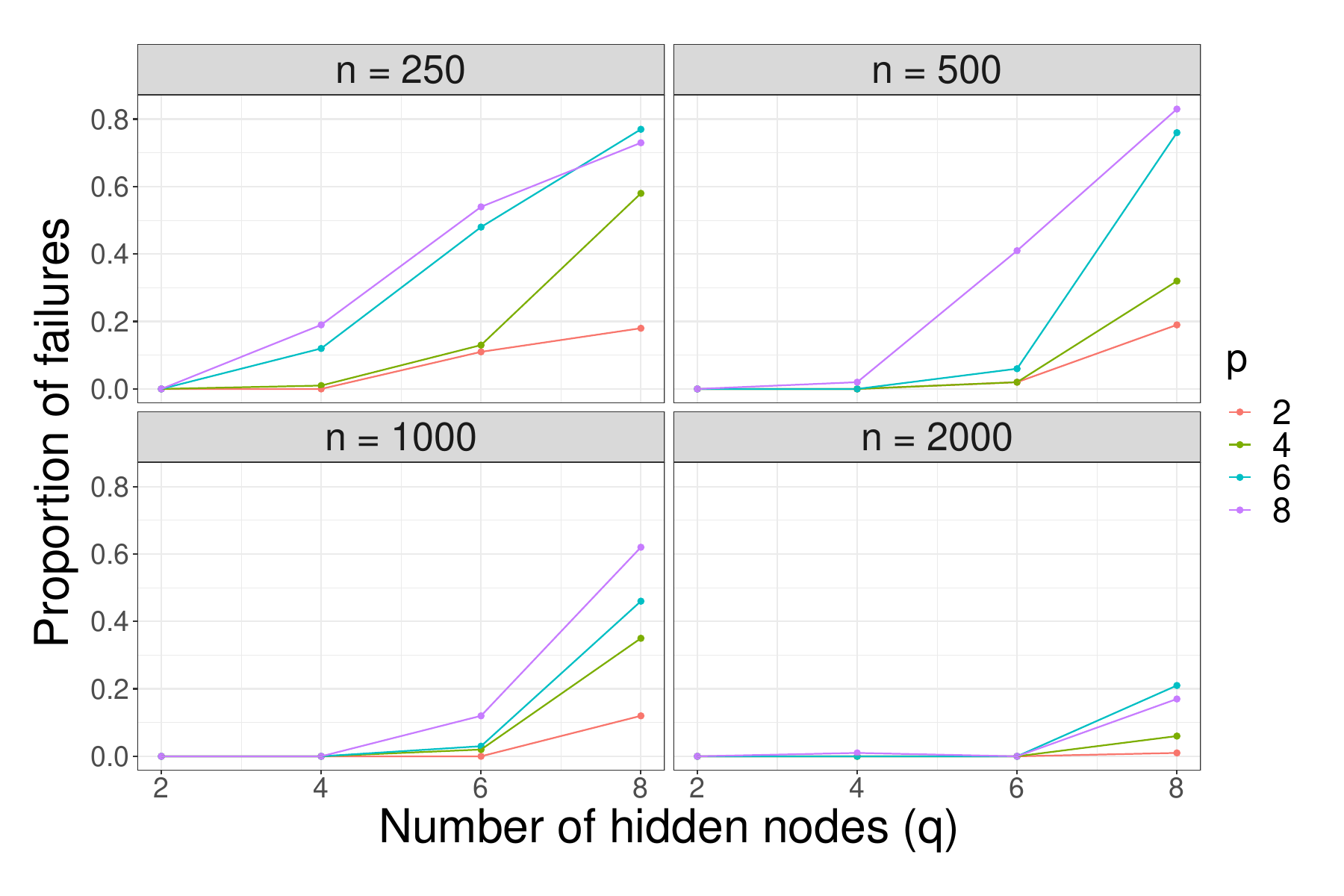}
    \caption{The proportion of replicates that the computation of EDF failed due to a non-invertible Hessian matrix.}
    \label{fig: sim_edf_failures}
\end{figure}

Aside from the reduced penalisation relative to $K$, it is also worth noting that GDF and EDF have other drawbacks.
First, GDF is quite computationally intensive to compute; a plot of the average time taken to compute GDF is given in Figure \ref{fig: sim_gdf_time}.
The computational times observed (which further increase with sample size and model complexity) render the use of GDF within model selection less feasible since a degrees-of-freedom value is needed at each step of the selection procedure.
As for the EDF computation, this requires the inversion of the Hessian matrix of the neural network parameters, which is not possible when there are redundancies present in the model; this is something that becomes more likely in more complex models.
Figure \ref{fig: sim_edf_failures} displays the proportion of replicates where the EDF could not be computed for each architecture and sample size.
It is clear that larger sample sizes are required for more complicated architectures in order to ensure stable computation of EDF.

\begin{figure}[t]
    \centering
    \includegraphics[width=\textwidth]{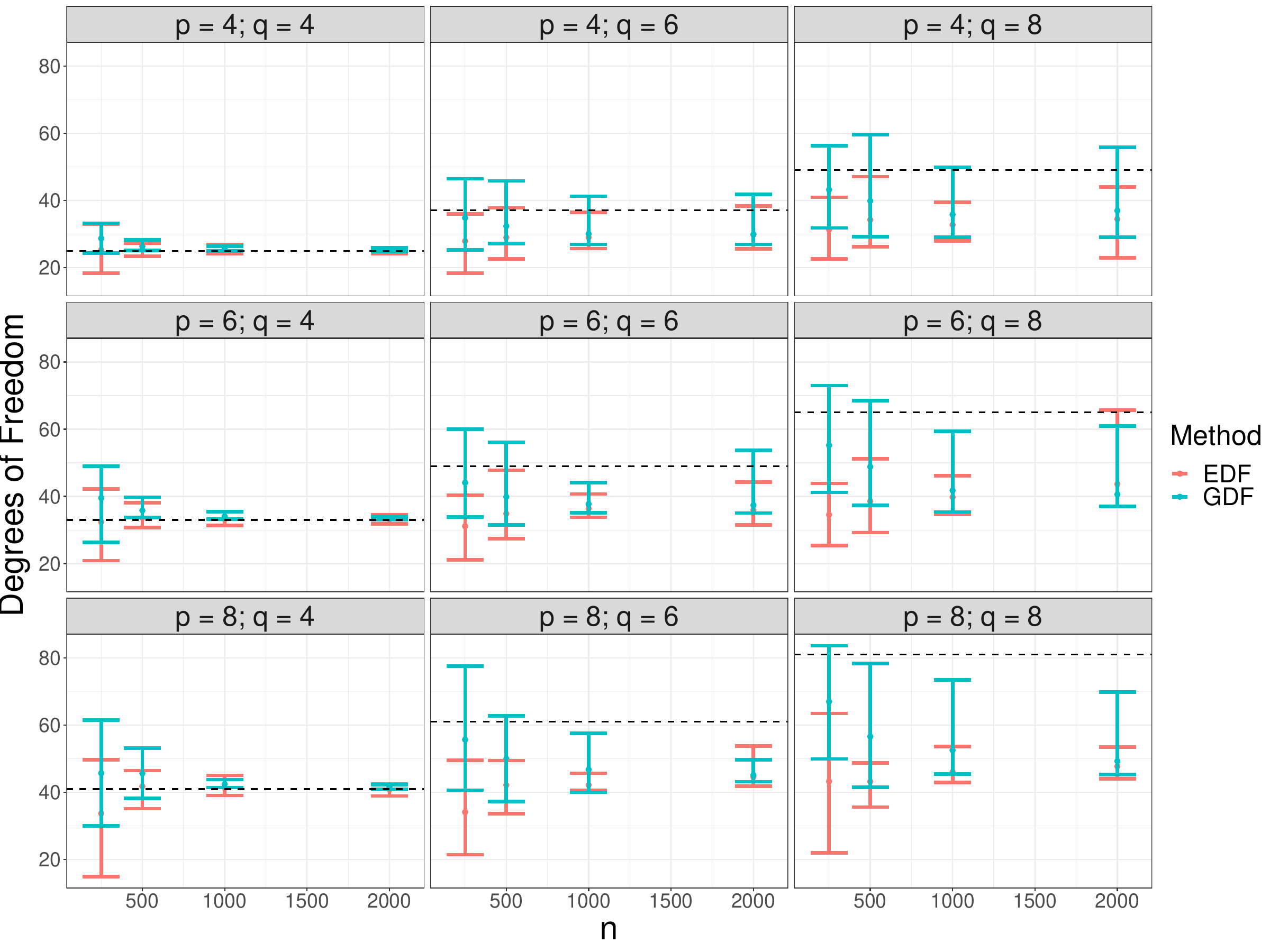}
    \caption{The average EDF and GDF values and their associated 2.5 and 97.5 quantiles versus sample size for different neural network architectures. The horizontal dashed line represents the number of parameters for each architecture.}
    \label{fig: sim_df_2}
\end{figure}

The second simulation study is similar to the first, but now the neural network architecture is misspecified.
The true data-generating model has $p = 4$ input nodes and $q = 4$ hidden nodes.
Neural networks of various architectures are fit to the data, with $p, q \in \{2, 4, 6, 8\}$.
The results are displayed in Figure \ref{fig: sim_df_2}.
For all models where the hidden layer is correctly specified ($q = 4$), the results are similar to Figure \ref{fig: sim_df_1}, i.e., the GDF and EDF approaches align with the number of parameters, $K$.
However, when the number of hidden nodes is incorrectly specified, EDF and GDF do not tend to the number of parameters.
In these scenarios, the degrees of freedom is lower than the number of parameters in the model; this suggests redundancies in the fitted model.

Finally, we compared all three approaches to defining the degrees of freedom within our proposed model selection procedure.
Due to the computational expense of GDF, a simpler simulation set up is used compared to the simulation study in the main paper.
Here, the true data-generating model is the same as the neural network used in Sections 4.1 and 4.2.
However, only two unimportant inputs and a $q_{\text{max}} = 5$ are considered.
The results of the model selection simulation are displayed in Table \ref{tab: df_selection}.

From the simulation results, it is clear that using GDF or EDF within the BIC penalty leads to reduced performance in the model selection procedure compared with using $K$.
This is likely due to the penalty being weaker when there is redundancy present (as seen in Figure \ref{fig: sim_df_2}), which leads to the selection of larger-than-required models (evidenced in the inflated false discovery rates in both the input and hidden nodes).
The median out-of-sample mean squared error (OOS) evaluated on a test set (20\% the size of the training set) is also reported.
The models selected using $K$ as the degrees-of-freedom term have better predictive performance than the models selected using the EDF and GDF approaches (albeit the OOS values appear to converge with the sample size).

\begin{table}[h!]\centering
\begin{threeparttable}
\caption{Simulation results: model selection metrics for different degrees-of-freedom approaches.}
\label{tab: df_selection} 
\begin{tabular}{ccccccccc}
  \toprule
       &           & \multicolumn{3}{c}{Input layer}  & \multicolumn{2}{c}{Hidden layer}  \\
  $n$  & Method & TNR & FDR & PI  & $\bar{q}$ (3) & PH &  OOS  & PT \\
   \cmidrule(lr){1-1} \cmidrule(lr){2-2}   \cmidrule(lr){3-5} \cmidrule(lr){6-7} \cmidrule(lr){8-9}

       & K     & 0.85 & 0.07  & 0.77  & 3.24 & 0.47 & 0.89 & 0.38 \\ 
  250  & GDF   & 0.56 & 0.20  & 0.29  & 4.93 & 0.05 & 1.10 & 0.00 \\
       & EDF    & 0.22 & 0.33 & 0.04  & 5.80 & 0.02 & 1.49 & 0.00 \\
       
   \cmidrule(lr){1-1} \cmidrule(lr){2-2}   \cmidrule(lr){3-5} \cmidrule(lr){6-7} \cmidrule(lr){8-9}

       & K      & 0.94 & 0.03 & 0.90  & 3.08 & 0.92 & 0.53 & 0.82 \\ 
  500  & GDF    & 0.74 & 0.12 & 0.55  & 4.66 & 0.15 & 0.60 & 0.10 \\ 
       & EDF    & 0.60 & 0.18 & 0.40  & 5.30 & 0.03 & 0.62 & 0.02 \\ 
       
   \cmidrule(lr){1-1} \cmidrule(lr){2-2}   \cmidrule(lr){3-5} \cmidrule(lr){6-7} \cmidrule(lr){8-9}
     
       & K      & 0.97 & 0.01 & 0.95  & 3.04 & 0.98 & 0.47 & 0.98 \\ 
 1000  & GDF    & 0.74 & 0.12 & 0.57  & 5.61 & 0.02 & 0.51 & 0.02 \\ 
       & EDF    & 0.74 & 0.12 & 0.59  & 4.78 & 0.02 & 0.50 & 0.01 \\ 
  \bottomrule
\end{tabular}
{\footnotesize\begin{tablenotes}[para, flushleft]
	\end{tablenotes}}
\end{threeparttable}
\end{table}


\section{Simulation Results with Correlated Data}\label{app: corr_data}
The performance of the proposed H-I-F approach is further investigated in the setting where there is correlation among the covariates.
Here, the data-generating process is the same as in the main paper, however, the data is generated such that the covariates are multivariate normal, i.e., $x_1, x_2, \dotsc, x_{13} \sim \text{MVN}$ wherein $\text{corr}(x_j, x_k) = 0.7^{|j-k|}$.
The non-zero covariates are $x_1, x_7, x_{13}$.
The results corresponding to Simulation 1 and 2 are shown in Tables \ref{tab: approaches_withcorr} and \ref{tab: objfun_withcorr}, respectively.

\begin{table}[ht!]\centering
\begin{threeparttable}
\caption{Simulation results: model selection metrics with correlated data for different model selection approaches.}
\label{tab: approaches_withcorr} 
\begin{tabular}{ccccccccc}
  \toprule
       &           & \multicolumn{3}{c}{Input layer}  & \multicolumn{2}{c}{Hidden layer}  \\
  $n$  & Method & TNR & FDR & PI & $\bar{q}$ (3) & PH &  PT \\
   \cmidrule(lr){1-1} \cmidrule(lr){2-2}   \cmidrule(lr){3-5} \cmidrule(lr){6-7} \cmidrule(lr){8-8}
       & H-I      & 0.82 & 0.38 & 0.63 & 2.89 & 0.52 & 0.38 \\
       & I-H      & 0.27 & 0.71 & 0.03 & 2.80 & 0.57 & 0.02 \\ 
  250  & H-I-F    & 0.84 & 0.35 & 0.60 & 3.07 & 0.82 & 0.53 \\
       & I-H-F    & 0.35 & 0.71 & 0.02 & 2.77 & 0.54 & 0.02 \\
       & F        & 0.68 & 0.52 & 0.35 & 7.52 & 0.24 & 0.18 \\
       
   \cmidrule(lr){1-1} \cmidrule(lr){2-2}   \cmidrule(lr){3-5} \cmidrule(lr){6-7} \cmidrule(lr){8-8}
    
       & H-I      & 0.96 & 0.13 & 0.90 & 3.23 & 0.76 & 0.73 \\ 
       & I-H      & 0.65 & 0.54 & 0.44 & 2.99 & 0.95 & 0.43 \\ 
  500  & H-I-F    & 0.97 & 0.10 & 0.91 & 3.06 & 0.94 & 0.86 \\ 
       & I-H-F    & 0.68 & 0.53 & 0.42 & 2.99 & 0.92 & 0.40 \\ 
       & F        & 0.97 & 0.10 & 0.88 & 3.15 & 0.92 & 0.83 \\ 
       
   \cmidrule(lr){1-1} \cmidrule(lr){2-2}   \cmidrule(lr){3-5} \cmidrule(lr){6-7} \cmidrule(lr){8-8}
     
       & H-I      & 1.00 & 0.00 & 0.99 & 3.00 & 1.00 & 0.98  \\
       & I-H      & 0.86 & 0.32 & 0.76 & 2.99 & 0.99 & 0.76 \\ 
 1000  & H-I-F    & 1.00 & 0.00 & 0.99 & 3.00 & 1.00 & 0.99 \\ 
       & I-H-F    & 0.88 & 0.29 & 0.78 & 3.00 & 0.98 & 0.77 \\ 
       & F        & 0.99 & 0.03 & 0.97 & 3.03 & 0.98 & 0.96 \\ 
  \bottomrule
\end{tabular}
{\footnotesize\begin{tablenotes}[para, flushleft]
	\end{tablenotes}}
\end{threeparttable}
\end{table}

\begin{table}[h!]\centering
\begin{threeparttable}
\caption{Simulation results: model selection metrics for the proposed approach (H-I-F) with correlated data for different objective functions.}
\label{tab: objfun_withcorr} 
\begin{tabular}{cccccccccc}
  \toprule
       &   &          \multicolumn{3}{c}{Input layer}  & \multicolumn{2}{c}{Hidden layer}  \\
  $n$  & Method & TNR & FDR & PI & $\bar{q}$ (3) & PH &  K (16) & OOS Test &  PT \\
   \cmidrule(lr){1-1} \cmidrule(lr){2-2} \cmidrule(lr){3-5}  \cmidrule(lr){6-7} \cmidrule(lr){8-10} 
         & AIC   & 0.12 & 0.75 & 0.00 & 12.17 & 0.00 & 157 & 2.17 &  0.00   \\
  250    & BIC   & 0.84 & 0.35 & 0.60 & 3.07  & 0.82 & 16  & 0.57 &  0.53   \\
         & OOS   & 0.41 & 0.66 & 0.02 & 4.12 & 0.35 & 39 & 0.87 & 0.00  \\
   \cmidrule(lr){1-1} \cmidrule(lr){2-2} \cmidrule(lr){3-5}  \cmidrule(lr){6-7} \cmidrule(lr){8-10} 
         & AIC   & 0.13 & 0.74 & 0.00 & 11.71 & 0.00 & 155 & 1.25 &  0.00   \\
  500    & BIC   & 0.97 & 0.10 & 0.91 & 3.06  & 0.94 & 16  & 0.61 &  0.86   \\
         & OOS   & 0.49 & 0.63 & 0.02 & 3.74 & 0.44 & 37 & 0.64 & 0.00  \\
   \cmidrule(lr){1-1} \cmidrule(lr){2-2} \cmidrule(lr){3-5}  \cmidrule(lr){6-7} \cmidrule(lr){8-10} 
         & AIC   & 0.14 & 0.74 & 0.00 & 11.77 & 0.00 & 155 & 0.83 &  0.00   \\
  1000   & BIC   & 1.00 & 0.00 & 0.99 & 3.00  & 1.00 & 16  & 0.51 &  0.99   \\
         & OOS   & 0.52 & 0.62 & 0.02 & 3.78 & 0.43 & 34 & 0.52 & 0.00  \\
  \bottomrule
\end{tabular}
{\footnotesize\begin{tablenotes}[para, flushleft]
	\item[]{}
	\end{tablenotes}}
\end{threeparttable}
\end{table}


\newpage
\section{Simulation: Number of Initialisations}\label{app: sim4}

\begin{figure}[h!]
    \centering
    \includegraphics[width=13cm]{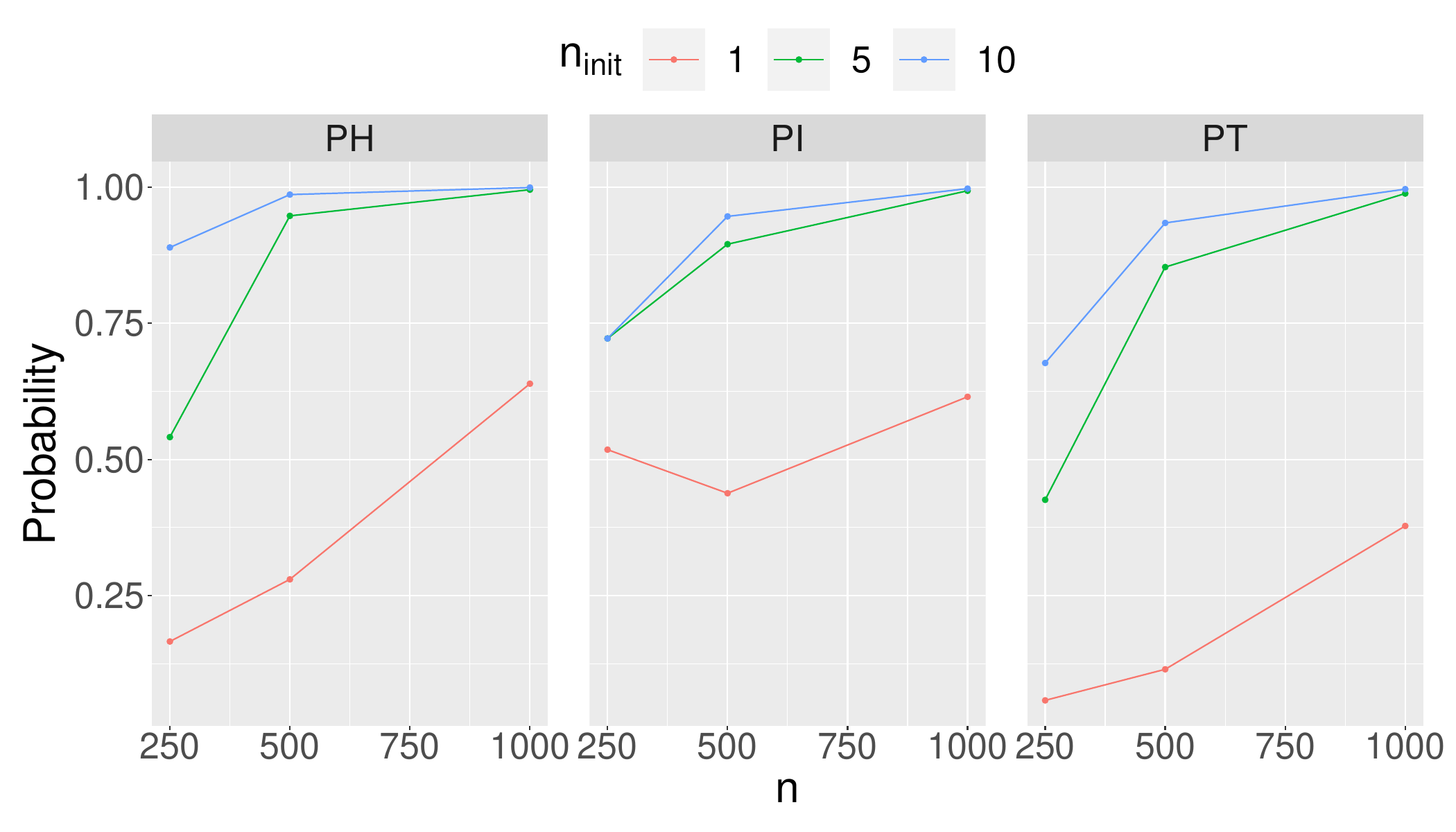}
    \caption{Simulation for number of initialisations: line plots for PH (the probability of choosing the correct number of hidden nodes), PI (the probability of choosing the correct set of inputs) and PT (the probability of choosing the overall true model) for different values of $n$ and $n_\text{init}$.}
    \label{fig: sim3_lineplot_PhPi}
\end{figure}

\begin{figure}[h!]
    \centering
    \includegraphics[width=13cm]{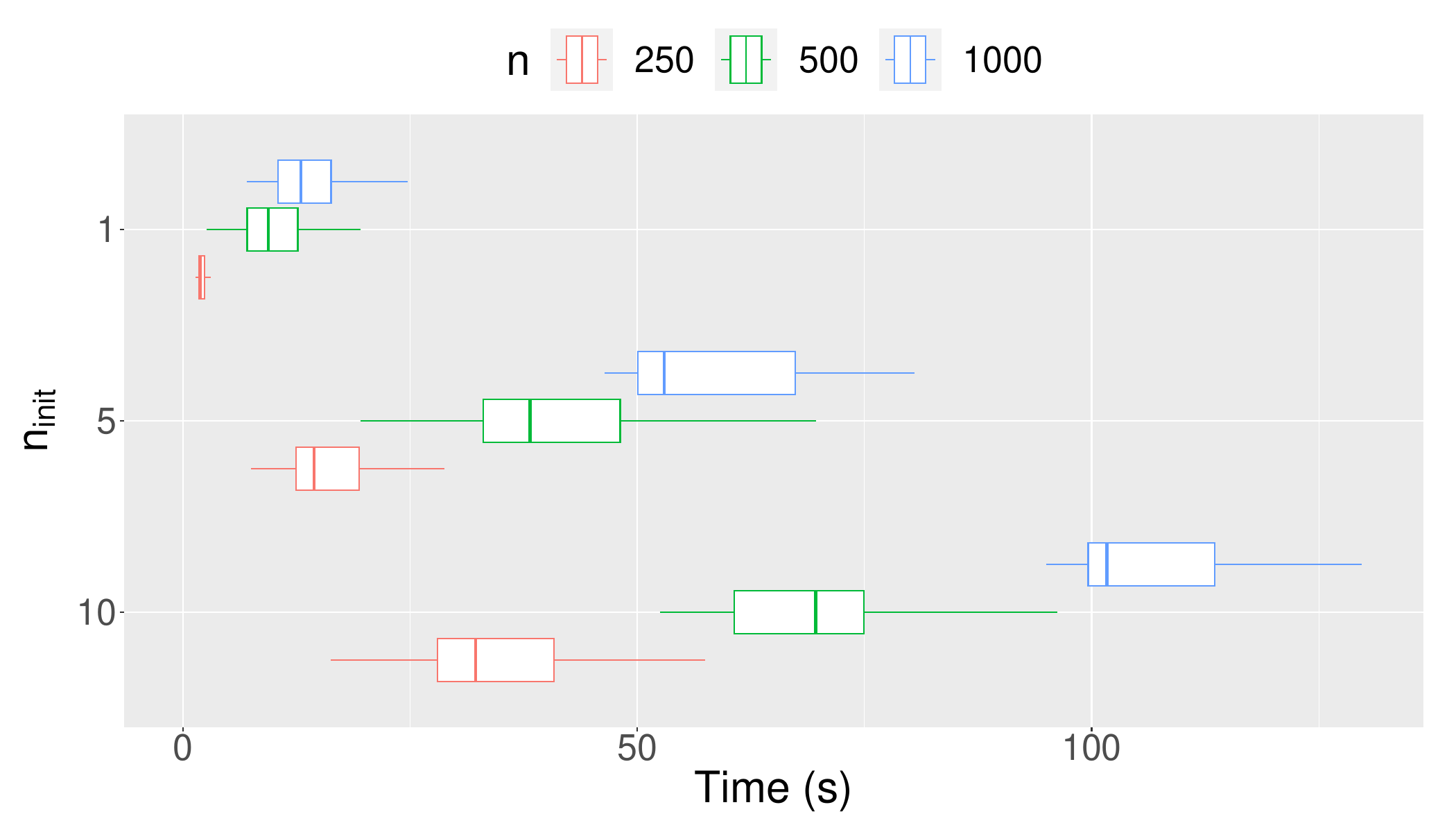}
    \caption{Simulation for number of initialisations: boxplots of computational time (s) for different values of $n$ and $n_\text{init}$.}
    \label{fig: sim3_boxplot_comptime}
\end{figure}

Since FNNs have a complex optimisation surface (the log-likelihood function), each model fit is supplied with $n_\text{init}$ random initial vectors with the aim of avoiding local maxima.
Of course, larger values of $n_\text{init}$ improve the chances of finding the global maximum but increase the computational expense.
In the previous simulations, we fixed $n_\text{init} = 5$, whereas, here, we vary it at $n_\text{init} \in \{1, 5, 10\}$ using the proposed H-I-F BIC-minimisation procedure.
Plots of the probability of choosing the correct number of hidden nodes (PH), the probability of choosing the correct set of inputs (PI), and the probability of choosing the overall true model (PT) for different values of $n$ and $n_\text{init}$ are shown in Figure~\ref{fig: sim3_lineplot_PhPi}.
Also, Figure~\ref{fig: sim3_boxplot_comptime} displays boxplots of the computational time for each scenario.
The corresponding table of simulation results is given in Table \ref{tab: n_init}.

From the plots, we can clearly see the trade-off between better model selection and worse computational efficiency as $n_\text{init}$ increases.
We would certainly recommend $n_\text{init} > 1$ initial vectors since the results are poor for $n_\text{init} = 1$.
Beyond this, the choice might be based on the computational constraints in a given practical setting, but we note, in particular, that larger values of $n_\text{init}$ are more important in smaller sample sizes.

\begin{table}\centering
\begin{threeparttable}
\caption{Simulation for number of initialisations: model selection metrics.}
\label{tab: n_init} 
\begin{tabular}{ccccccccc}
  \toprule
       & &           & \multicolumn{3}{c}{Input layer}  & \multicolumn{3}{c}{Hidden layer}  \\
  $n$  & $n_{\text{init}}$ & Time (s) & TNR & FDR & PI & $\bar{q}$ (3) & PH &  PT \\
   \cmidrule(lr){1-1} \cmidrule(lr){2-2}  \cmidrule(lr){3-3} \cmidrule(lr){4-6} \cmidrule(lr){7-8} \cmidrule(lr){9-9}

       & 1  & \textbf{1.90}  & 0.80 & 0.23 & 0.52  & 2.20 & 0.17 & 0.06 \\
   250 & 5  & 14.42 & 0.87 & 0.15 & \textbf{0.72} & 2.66 & 0.54 & 0.43 \\
       & 10 & 32.30 & \textbf{0.89} & \textbf{0.14} & \textbf{0.72}  & \textbf{3.05} & \textbf{0.89} & \textbf{0.68} \\

   \cmidrule(lr){1-1} \cmidrule(lr){2-2}  \cmidrule(lr){3-3} \cmidrule(lr){4-6} \cmidrule(lr){7-8} \cmidrule(lr){9-9}

        & 1  &  \textbf{9.37} & 0.73 & 0.31 & 0.44  & 3.82 & 0.28 & 0.12 \\
   500  & 5  & 38.18 & 0.96 & 0.05 & 0.90  & 3.05 & 0.95 & 0.85 \\
        & 10 & 70.69 & \textbf{0.98} & \textbf{0.03} & \textbf{0.95}  & \textbf{3.01} & \textbf{0.99} & \textbf{0.93} \\

   \cmidrule(lr){1-1} \cmidrule(lr){2-2}  \cmidrule(lr){3-3} \cmidrule(lr){4-6} \cmidrule(lr){7-8} \cmidrule(lr){9-9}

         & 1  &  \textbf{12.98} & 0.89 & 0.16 & 0.62  & 3.40 & 0.64 & 0.38 \\
   1000  & 5  & 54.39 & \textbf{1.00} & \textbf{0.00}  & 0.99  & \textbf{3.00} & \textbf{1.00} & 0.99 \\
         & 10 & 117.32 & \textbf{1.00} & \textbf{0.00} & \textbf{1.00} & \textbf{3.00} & \textbf{1.00} & \textbf{1.00} \\

  \bottomrule
\end{tabular}
{\footnotesize\begin{tablenotes}[para, flushleft]
	\item[]{Time (s), median time to completion in seconds (carried out on an Intel\textsuperscript{\textregistered} Core\textsuperscript{\texttrademark}  i5-10210U Processor).
		Best values for a given sample size are highlighted in \textbf{bold}.}
	\end{tablenotes}}
\end{threeparttable}
\end{table}


\clearpage

\section{Relationship Between the Structure of the Input and Hidden Layer \label{app: rel_structure}}
In Simulation 1 (Section 4.1), there appears to be a relationship between the structure of one layer on the probability of selecting the correct structure for the other layer.
For example, when $n = 500$, performing input-layer selection after first performing hidden-layer selection (H-I) results in a much higher probability of selecting the correct set of input nodes ($\text{PI} = 0.83$) in comparison to performing input-layer selection first (I-H) ($\text{PI} = 0.43$).
This suggests that input-layer selection is improved when the hidden layer is closer to its correct structure.
In order to investigate this further, we performed two simulation studies: one to explore the relationship between the input-layer structure and the probability of selecting the correct number of hidden nodes, and another to explore the relationship between the hidden-layer structure and the probability of selecting the correct set of input nodes.
In each simulation study, the response is generated from an FNN with known ``true" architecture containing five important inputs, $x_1, x_2, \dotsc, x_5$, and $q = 5$ hidden nodes.
Within each simulation, every scenario is implemented for 1,000 replicates, using a sample size of $n = 500$.

\begin{figure}[h!]
    \centering
    \includegraphics[width=0.64
    \textwidth]{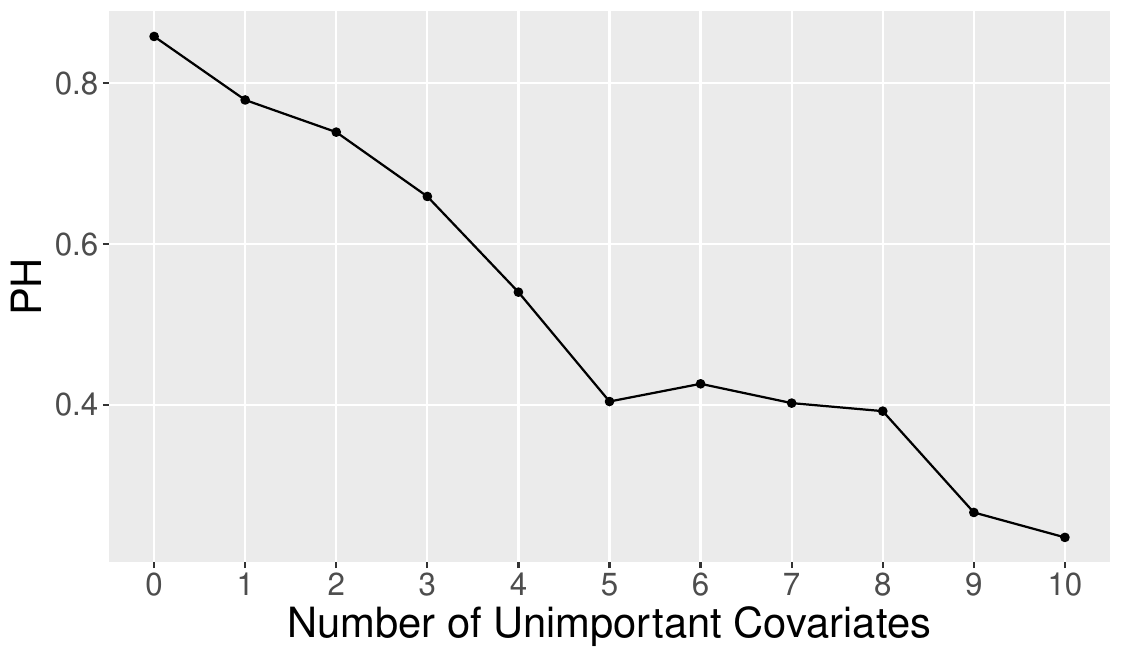}
    \caption{Number of unimportant covariates versus the probability of selecting the correct hidden-layer structure.}
    \label{fig: PH_vs_p}
\end{figure}

The aim of the first simulation study is to investigate the effect of adding additional unimportant covariates to the set of important covariates on hidden-node selection.
All important covariates remain in the data, while the number of unimportant inputs, $n_{\text{unimp}}$, is varied from zero up to ten.
The hidden-node selection step (Algorithm~2) is implemented for each replicate, with a maximum of 10 hidden nodes being considered.
The probability of choosing the correct number of hidden nodes (PH) is then calculated.
The results are displayed in Table~\ref{tab: PH vs p} and Figure~\ref{fig: PH_vs_p}.
It is clear that the more unimportant covariates that are included in the data, the lower the probability in recovering the correct hidden-layer structure.

\begin{table}[h]
\centering
\begin{threeparttable}
\caption{Probability of selecting the correct hidden-layer structure.}
\label{tab: PH vs p} 
\begin{tabular}{lccccccccccc}
  \toprule
  $n_{\text{unimp}}$ & 0  & 1  &  2  &  3  & 4  &  5  &  6  &  7  &  8  &  9  & 10  \\ 
  \cmidrule(l){2-12}
  PH & 0.86 & 0.78 & 0.74 & 0.66& 0.54 & 0.40 & 0.43 & 0.40 & 0.39 & 0.27 & 0.23 \\ 
  \bottomrule
\end{tabular}
{\footnotesize\begin{tablenotes}[para, flushleft]
	\item[]{$n_{\text{unimp}}$, the number of unimportant covariates; PH, the probability of selecting the correct number of hidden nodes.}
	\end{tablenotes}}
\end{threeparttable}
\end{table}

\begin{figure}[h!]
    \centering
    \includegraphics[width=0.64
    \textwidth]{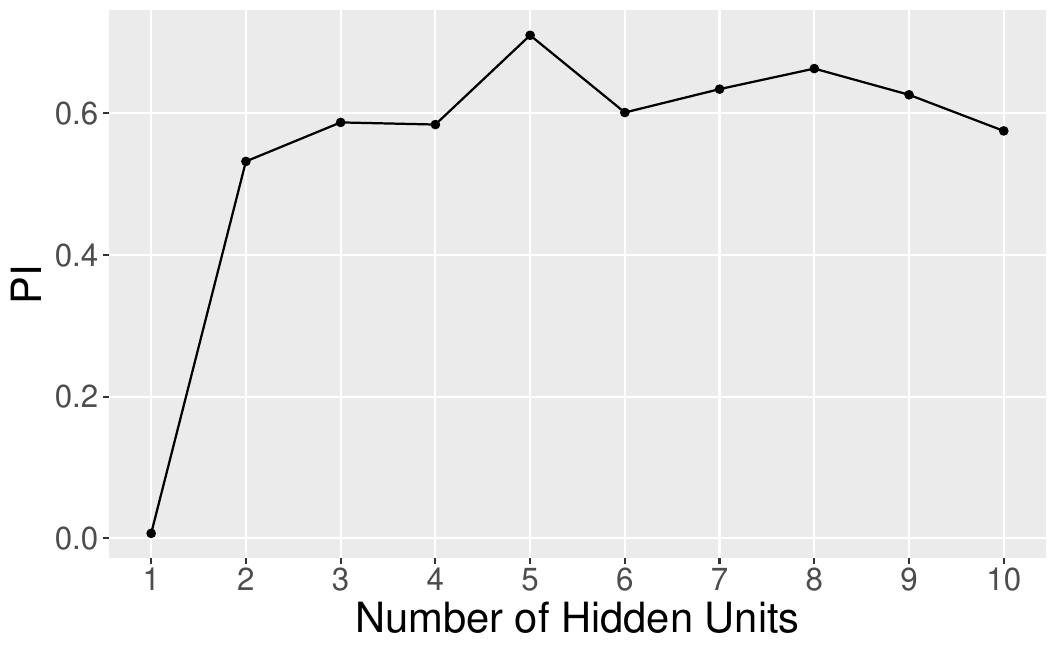}
    \caption{Number of hidden nodes versus the probability of selecting the correct input-layer structure.}
    \label{fig: PI_vs_q}
\end{figure}

The second simulation study aims to investigate the effect of the number of hidden nodes on input-node selection.
Ten additional unimportant inputs, $x_6, x_7, \dotsc, x_{15}$, are added to the data.
The number of hidden nodes is varied from one up to ten.
The input-node selection step (Algorithm~3) is implemented for each replicate, and the probability of choosing the correct set of input nodes (PI) is calculated.
The results are displayed in Table~\ref{tab: PI vs q} and Figure~\ref{fig: PI_vs_q}.
We find that the closer the hidden layer is to having the correct number of hidden nodes, the greater the probability of recovering the correct set of input nodes.
Both simulation studies verify that model selection for one layer is dependent on the structure of the other layer, and, hence, justifies the use of a fine-tuning phase after performing input- and hidden-node selection.

\begin{table}[h]
\centering
\begin{threeparttable}
\caption{Probability of selecting the correct input-layer structure.}
\label{tab: PI vs q} 
\begin{tabular}{lcccccccccc}
  \toprule
  $q$ & 1  &  2  &  3  & 4  &  5  &  6  &  7  &  8  &  9  & 10  \\ 
  \cmidrule(l){2-11}
  PI & 0.01 & 0.53 & 0.59 & 0.58 & 0.71 & 0.60 & 0.63 & 0.66 & 0.63 & 0.58 \\ 
  \bottomrule
\end{tabular}
{\footnotesize\begin{tablenotes}[para, flushleft]
	\item[]{$q$, the number of hidden nodes; PI, the probability of selecting the correct set of input nodes.}
	\end{tablenotes}}
\end{threeparttable}
\end{table}

\section{Simulation 1: Boxplots of Computational Time for Each Model Selection Method}\label{app: sim1_comptime}

This section contains the boxplots associated with the computational time for the different model selection approaches, and corresponds to Table 1 in the main paper.

\begin{figure}[h!]
    \centering
    \includegraphics[width=14cm]{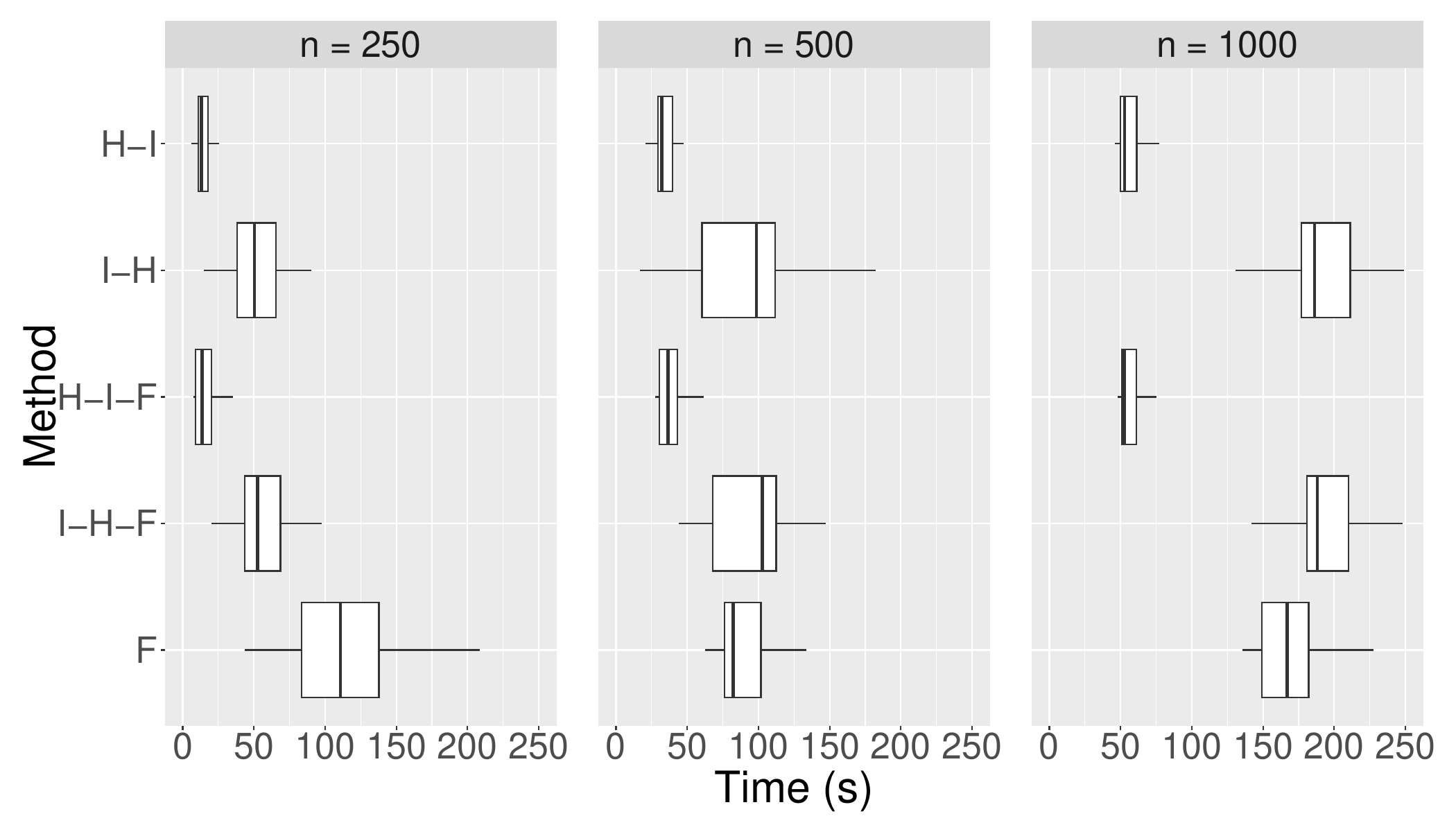}
    \caption{Simulation 1: boxplots of computational time (s) for each model selection approach and different values of $n$.}
    \label{fig: sim1_boxplot_comptime}
\end{figure}

\section{Simulation 1: Combined Approach \label{app: combapproach}}
Since H-I yields a higher probability of recovering the \emph{input} layer than I-H, and I-H yields a higher probability of recovering the \emph{hidden} layer than H-I, we have also considered the following approach: run both H-I and I-H independently, then take the input layer from H-I and the hidden layer from I-H to form the model; we refer to this as [H-$\text{I}^*$]-[I-$\text{H}^*$], where an asterisk denotes the layer being selected.
We also investigate the aforementioned approach but followed by a fine-tuning phase, which we refer to as [H-$\text{I}^*$]-[I-$\text{H}^*$]-F. 
The simulation results for these procedures are given in Table~\ref{tab: combapproach}.

We find that the [H-$\text{I}^*$]-[I-$\text{H}^*$] approach outperforms both the H-I and the I-H approaches, but is more computationally expensive.
The addition of a fine-tuning phase improves the model selection performance further, but the proposed H-I-F approach has very similar performance while being much less computationally demanding. 

\begin{table}\centering
\begin{threeparttable}
\caption{Model selection metrics for combined approaches.}
\label{tab: combapproach} 
\begin{tabular}{ccccccccc}
  \toprule
       & &           & \multicolumn{3}{c}{Input layer}  & \multicolumn{2}{c}{Hidden layer}  \\
  $n$  & Method & Time (s) & TNR & FDR & PI & $\bar{q}$ (3) & PH &  PT \\
   \cmidrule(lr){1-1} \cmidrule(lr){2-2}  \cmidrule(lr){3-3} \cmidrule(lr){4-6} \cmidrule(lr){7-8} \cmidrule(lr){9-9}

   250  & [H-$\text{I}^*$]-[I-$\text{H}^*$] & 44.96  & 0.83 & 0.20 & 0.61 & 2.89 & 0.42 & 0.25 \\
   500  & [H-$\text{I}^*$]-[I-$\text{H}^*$] & 131.50 & 0.90 & 0.11 & 0.79  & 3.19 & 0.81 & 0.64 \\
   1000 & [H-$\text{I}^*$]-[I-$\text{H}^*$] & 238.15 & 1.00 & 0.00 & 0.98  & 3.00 & 1.00 & 0.98 \\

   \cmidrule(lr){1-1} \cmidrule(lr){2-2}  \cmidrule(lr){3-3} \cmidrule(lr){4-6} \cmidrule(lr){7-8} \cmidrule(lr){9-9}

   250  & [H-$\text{I}^*$]-[I-$\text{H}^*$]-F &  45.72 & 0.85 & 0.18 & 0.65  & 2.81 & 0.59 & 0.46 \\
   500  & [H-$\text{I}^*$]-[I-$\text{H}^*$]-F & 134.73 & 0.93 & 0.08  & 0.85  & 3.04 & 0.96 & 0.82 \\
   1000 & [H-$\text{I}^*$]-[I-$\text{H}^*$]-F & 240.06 & 1.00 & 0.01 & 0.99  & 3.00 & 1.00 & 0.98 \\

  \bottomrule
\end{tabular}
{\footnotesize\begin{tablenotes}[para, flushleft]
	\item[]{Time (s), median time to completion in seconds (carried out on an Intel\textsuperscript{\textregistered} Core\textsuperscript{\texttrademark}  i5-10210U Processor).}
	\end{tablenotes}}
\end{threeparttable}
\end{table}



\section{Simulation 2: Boxplots of TNR and $q$ for Different Model Selection Objective Functions}\label{app: sim2_boxplots}

This section contains the boxplots associated with TNR and $q$ corresponding to Table 2 in Section 4.2 of the main paper.

\begin{figure}[b!]
    \centering
    \includegraphics[width=\textwidth]{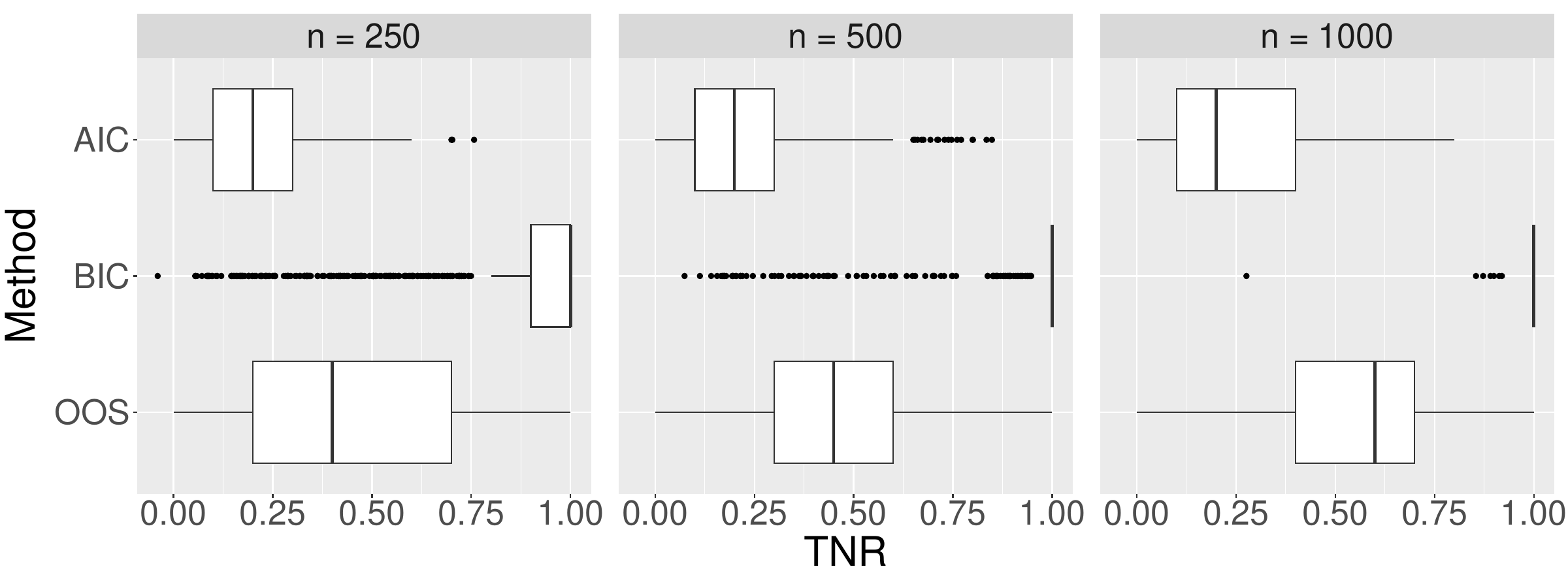}
    \caption{Simulation 2: boxplots for TNR (the true negative rate for the input variables) for the models selected by each objective function.}
    \label{fig: sim2_boxplot_C}
\end{figure}

\begin{figure}[t]
    \centering
            \includegraphics[width=\textwidth]{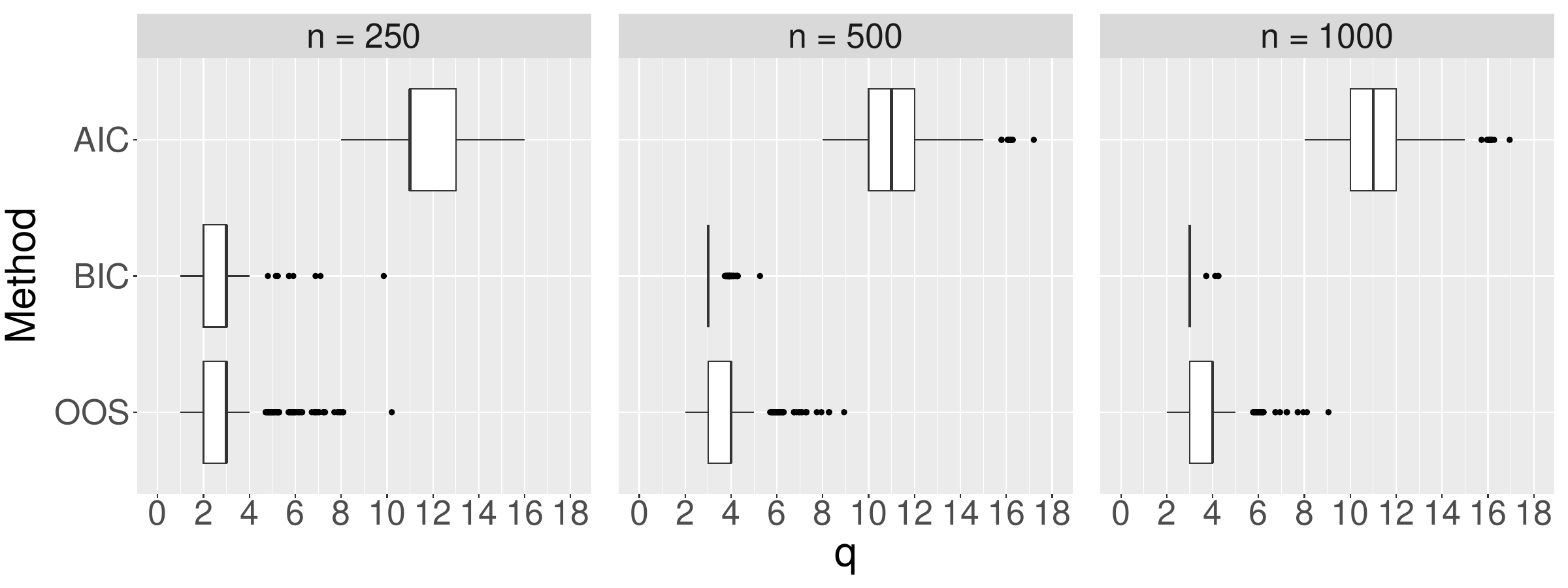}
    \caption{Simulation 2: boxplots for q (the number of hidden nodes selected) for the models selected by each objective function.}
    \label{fig: sim2_boxplot_q}
\end{figure}

\newpage
\section{Comparison of models selected with the full model trained using weight decay and early stopping}\label{app: wd_and_es}
\begin{figure}[h!]
    \centering
    \includegraphics[width=16cm]{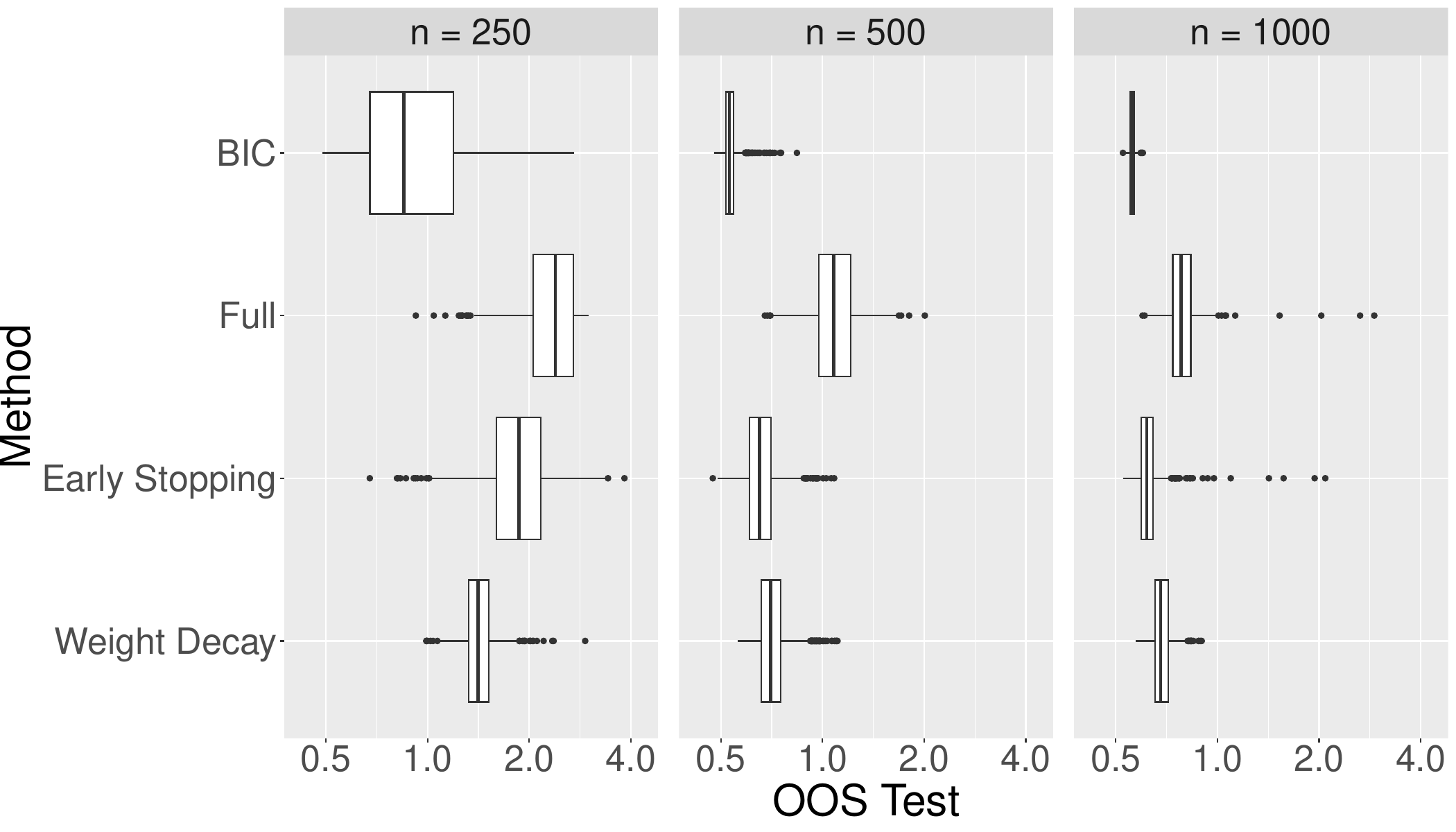}
    \caption{Boxplots of out-of-sample mean squared error evaluated on a test data set (on a logarithmic scale) for the models selected using the proposed approach, the full model, the full model with early stopping, and the full model with weight decay.}
    \label{fig: boxplots_oos_test}
\end{figure}

Over-parameterised neural networks can often suffer from issues of overfitting.
Therefore, it is of interest to compare the out-of-sample performance of the models selected using our stepwise BIC procedure with the out-of-sample performance of the full model trained with common approaches that deal with overfitting.
We investigate two popular approaches to overfitting: the use of a weight decay penalty, and early stopping.
The weight decay penalty and the stopping point are both chosen based on the performance of the model on an additional validation data set (which is 20\% the size of the training data set).
Boxplots of the out-of-sample performance on the test data set for the BIC-selected model, the full model, and the models trained with weight decay and early stopping are displayed in Figure \ref{fig: boxplots_oos_test}.

It is clear that both weight decay and early stopping improve the out-of-sample performance of the full model. 
However, the more parsimonious models that are selected via the proposed approach outperform both the commonly used weight decay and early stopping strategies.

\newpage
\section{$\Delta\text{BIC}$ and simple covariate effects}\label{app: delta_bic}
Simple measures of covariate importance and effects are used in Section 5 to accompany the model selection procedure for the data application.
For the covariates that are selected, their relative importance is estimated using BIC differences ($\Delta\text{BIC}$), which is given by $\Delta\text{BIC}_j = \text{BIC}_j - \text{BIC}_{\text{min}}$,
where $\text{BIC}_{\text{min}}$ is the BIC for the selected model (i.e., it is the model with the minimum BIC found by our algorithm) and $\text{BIC}_j$ is the BIC for the model with the same hidden-layer structure as this selected model but with covariate $j$ removed; hence, the more important covariate $j$ is, the larger the corresponding $\Delta\text{BIC}_j$ value as its removal would lead to an increased BIC$_j$ compared to BIC$_{\text{min}}$.
In addition to covariate importance, simple covariate effects ($\hat \tau_j$) are constructed by splitting the data into two groups based on whether or not the value of $j$th covariate for the $i$th individual, $x_{ji}$, is above or below the empirical median value, $m_j = \text{median}(x_{j1},x_{j2},\dotsc, x_{jn})$, and computing the difference in the average predicted response for these two groups.
The corresponding equation is
\begin{equation*}
    \hat{\tau}_j = E[\text{NN}(X) \mid X_{(j)} > m_j] - E[\text{NN}(X) \mid X_{(j)} < m_j],
\end{equation*}
where $\text{NN}(X)$ is the output of the neural network for input covariate vector $X$ (see Equation~2.1), $X_{(j)}$ denotes the $j$th covariate, and we take the conditional expected value with respect to the empirical distribution of covariates in the dataset, i.e.,
\begin{equation*}
    E[\text{NN}(X)\mid X_{(j)} > m_j] = \frac{1}{\text{card}(i\mid x_{ji} > m_j)}\sum_{i\mid x_{ji} > m_j} \text{NN}(x_i)
\end{equation*}
 where $\text{card}(\cdot)$ is the cardinality operator.
Although this metric is a simplification of the (potentially non-linear) effects captured by the neural network, and our focus is on model selection, it is nevertheless a useful supplement to our approach that provides a high-level overview of the estimated covariate effects.
\end{document}